\documentclass[useAMS,usenatbib,epsf]{mn2e}
\begin{document}
\input{epsf}
\title[Imaging the transition between pre-PNe and PNe]{Imaging the transition between pre-planetary and planetary nebulae: Integral Field 
Spectroscopy of hot post-AGB stars with NIFS}

\author[T.M. Gledhill et al.]
        {T.M.~Gledhill$^{1}$\thanks{email: {\tt t.gledhill@herts.ac.uk}},
         K.P.~Forde$^{1}$\\
         $^{1}$Science and Technology 
         Research Institute, University of Hertfordshire, 
         College Lane, Hatfield AL10 9AB, UK \\
         }
\maketitle

\begin{abstract}

We present 2--2.4~$\umu$m integral field spectroscopy of a sample of
hot post-AGB stars with early-B spectral types, using the NIFS
instrument on Gemini North. These stars are just beginning to ionize
their immediate environments and turn into planetary nebulae (PNe).
We use molecular hydrogen emission lines together with hydrogen
and helium recombination lines to explore the distribution of
molecular and atomic gas and the extent of the developing ionized
region. We see a range of evolutionary stages: IRAS 18062+2410 and
IRAS~18379-1707 have recently developed compact and unresolved regions
of photoionized H within axisymmetric molecular envelopes, with the
former object increasing its Br$\gamma$ flux by a factor of 5.3 in 14
years; IRAS~22023+5249 and IRAS~20462+3416 have extended Br$\gamma$
nebulae and in the latter object only weak H$_2$ emission remains;
IRAS~19336-0400 is at a more advanced stage of PN formation where
H$_2$ is mostly dissociated and we see structure in both the H and He
recombination line nebulae. IRAS~19200+3457 is the only object not to
show the He{\sevensize~I} line at $2.058~\umu$m and is probably the
least evolved object in our sample; the H$_2$ emission forms a 
ring around the star and we suggest that this object may be a
rare example of a ``round'' pre-PN in transition to a ``round'' PN.

\end{abstract}


\section{Introduction}

In the early stage of post-asymptotic giant branch (post-AGB)
evolution, stars have spectral types typically from K to F and the
envelope material ejected on the AGB is seen in scattered and thermal
emission from dust and in transitions from molecules such as CO and
H$_2$. These envelopes form pre-planetary nebulae (pre-PNe). Imaging
studies of pre-PNe at optical and infrared (IR) wavelengths have
resolved complex structures, with a wide range of bipolar, multipolar
and point-symmetric morphologies seen in dust-scattered light
(e.g. Ueta, Meixner \& Bobrowsky 2000; Gledhill et al. 2001; Gledhill
2005; Ueta et al. 2005; Sahai et al. 2007; Ueta et al. 2007;
Si\'{o}dmiak et al. 2008) as well as in the thermal emission from dust
(e.g. Meixner et al. 1999; Lagadec et al. 2011). Although the exact
mechanism by which the shaping occurs is still debated, it seems
likely that interaction between the mass-losing star and a nearby
companion is involved (review by De Marco 2009). It is also apparent
from the structures observed and their presence around later spectral
type stars, that the shaping process must commence close to the end of
the AGB (Ueta et al. 2000; Meixner et al 1999).

As the post-AGB star evolves to hotter temperatures and earlier
spectral types it will begin to photodissociate and ionize the pre-PN
material. The signatures of ionization, in the form of hydrogen
recombination lines, become detectable for early-B spectral types and
if evolution to this point occurs before the ejected material
disperses into the interstellar medium, then a PN can form.  Imaging
surveys of young PNe in optical emission lines with {\em HST} show
strong morphological similarities with the structure seen in the
pre-PN phase (Sahai \& Trauger 1998; Sahai, Morris \& Villar
2011). These surveys select young PNe based on a O[III]/H$\alpha$ flux
ratio of less than one, corresponding to central stars with $T_{\rm
  eff}$ in the range $25~000-40~000$~K.  The H ionization front in
these PNe will typically have expanded to encompass all of the material
shaped during the pre-PN phase.

\begin{table*}
\caption{Observation log, including date of observation, exposure and
  integration times in seconds, the full width at half maximum (FWHM)
  of the standard star as an indication of the adaptive optics
  correction, and the air mass range.}
\begin{tabular}{lccccccc}
IRAS ID    & Other Designation  & Obs. Date & Exp. (s) & Integ. (s) & Standard & FWHM (arcsec) & AM \\
\hline 
18062+2410 & LSE 162, V886 He   & 20070505  & 180      & 1440  & HIP~83274 & $0.16\times 0.14$ & $1.54-1.01$ \\
18379-1707 & LS 5112, PM 2-37   & 20070508  & 200      & 1600  & HIP~96674 & $0.14\times 0.13$ & $1.52-1.25$\\
19200+3457 & LS II +341, ALS 10266&20070506 & 150      & 1200  & HIP~95414 & $0.18\times 0.13$ & $1.08-1.04$\\
19336-0400 & PN G034.5-11.7     & 20070522  & 400      & 1200  & HIP~102221 & $0.13\times 0.12$ &$1.11-1.15$ \\
20462+3416 & LS II +3426, PN G076.6-05.7&20070528&150  &600    & HIP~106674 & $0.12\times 0.11$ &$1.03-1.04$\\
22023+5249 & LS III +52 24, PN G099.3-01.9&20070530&85&1105    & HIP~103685 &$0.13\times 0.12$  &$1.77-1.31$\\
\hline \\
\end{tabular}
\end{table*}
In the intervening transition stage between pre-PN and PN, the
ionization front is beginning to break out of the central region, but
most of the nebula material remains molecular. A population of objects
thought to be in this transition stage has been identified
(e.g. Parthasarathy \& Pottasch 1989; Parthasarathy, Vijapurkar,
Drilling 2000) and termed 'hot post-AGB stars', but few imaging
studies have been made so far. In these cases we cannot rely
exclusively on recombination lines to reveal the nebula structure, but
must also include tracers of the molecular material. We find that the
$K$-band between 2 and 2.4~$\umu$m provides a convenient wavelength
range for the investigation of these objects, containing a broad range
(in excitation) of ro-vibration transitions of H$_2$ as well as the
Br$\gamma$ and He~{\sevensize I}~$2.058~\umu$m recombination lines.
The near-IR is also suited to studies of very early PNe in which the
ionized region is expected to be compact and may therefore be
optically obscured by circumstellar dust.

The purpose of this study is to image both the molecular pre-PN
material and the developing ionized regions in objects that are in
transition between the pre-PN and PN phases. The formation and
structure of PNe is expected to depend specifically on how the
ionization front develops and propagates through the circumstellar
material shaped during the pre-PN phase. We use the Near-infrared
Integral Field Spectrometer (NIFS) on the 8.2-m Gemini North telescope
to investigate the spatial and spectral distribution of emission from
neutral and ionized gas in a small sample of objects which are at the
point of becoming PNe. Combined with adaptive optics (AO) correction,
angular resolutions of up to 0.1 arcsec can be achieved in the
$K$-band, with a spectral resolution of about $5~000$. We observe six
post-AGB stars with B1I spectral types, which belong to the class of
`hot post-AGB stars'. H$_2$ emission has previously been detected
spectroscopically in all six objects by Kelly \& Hrivnak (2005;
hereafter KH05).  Radio emission from the ionized region is detected
in four objects (Cerrigone et al. 2011).

\section{Observations and data reduction}

Observations were made in May 2007 using NIFS on the 8.2-m Gemini
North Telescope (McGregor et al. 2002), programme number
GN-2007A-Q-79.  The K-G5605 grating resulted in a usable wavelength
range of $2.013-2.433~\umu$m in the reduced data, with each spectral
channel being $2.13\times 10^{-4}~\umu$m wide, corresponding to
approximately 30~km~s$^{-1}$ in the $K$-band. The full-width at
half-maximum (FWHM) in the spectral direction is approximately two
channels, as measured from the arc lines. AO correction was achieved
with the ALTAIR facility, using natural guide stars. Typically 4
pointings were observed for each target, so that the observed area on
the sky is slightly larger than the $3\times 3$~arcsec field of view
of NIFS.

Data reduction was accomplished using the Gemini {\sevensize IRAF}
package, driven in a semi-automated fashion using modified scripts
from the Gemini NIFS web pages. The reduction stages included dark and
flat-field correction, wavelength calibration, telluric correction and
image combination, to form a final 3-dimensional data cube. The cube
was then flux calibrated using an appropriate standard star. The
reduction process results in a pixel size of $0.05 \times
0.05$~arcsec. The point spread function (PSF) size in the reduced
images, as measured from the standard star, is given in Table~1 for
each target, along with other observational details. The FWHM does not
vary by more than 0.02 arcsec across the NIFS wavelength range.

Visualization of the data cubes (along with extraction of images and
spectra) used the GAIA image display tool and other applications
from the {\sevensize STARLINK} software collection. $K$-band spectra 
were extracted by integrating within a circular aperture centred on each
object, and are shown in Fig.~1 along with the aperture size.

To facilitate further analysis, the data cubes for each object were
split in the wavelength dimension into smaller cubes, each containing
an emission line (or group of lines if close in wavelength) and
surrounding continuum. A linear fit to the wavelength-dependence of
the continuum was made for each spatial pixel to produce a cube
containing the estimated continuum emission. This was then subtracted
from the object data cube to produce a continuum-subtracted cube
containing only the line emission. These cubes were used to form
images of emission lines and to measure the line flux.

\begin{table*}
\caption{Line fluxes ($\times 10^{-17}$~W~m$^{-2}$ and $\pm 0.01$) and line ratios for key $K$-band 
emission lines.}
\begin{tabular}{lcccccccccc}
IRAS ID &1-0~S(1)&2-1~S(1)&3-2~S(3)&He~{\sevensize I}&Br$\gamma$&Mg~{\sevensize II}&$\frac{1-0~{\rm S(1)}}{2-1~{\rm S(1)}}$&$\frac{1-0~{\rm S(1)}}{3-2~{\rm S(3)}}$&$\frac{1-0~{\rm S(1)}}{{\rm Br}\gamma}$&$\frac{\rm He~{\sevensize I}}{{\rm Br}\gamma}$ \\
\hline 
18062+2410& 3.09 & 0.51 & 0.16 & 1.06 & 4.35 & 0.22,$a$ & $6.1\pm 0.1$ & $19\pm 1$ & $0.71\pm 0.01$ & $0.24\pm 0.01$ \\
18379-1707& 6.26 & 1.16 & 0.50 & 1.32 & 0.67 & 0.21,0.09 & $5.40\pm 0.05$ & $12.5\pm 0.3$ & $10.5\pm 0.2$ & $1.99\pm 0.03$ \\
19200+3457& 6.53 & 1.16 & 0.47 &nd& 1.41 & 0.13,0.07& $5.63\pm 0.05$ & $13.9\pm 0.3$ & $4.63\pm 0.03$ & -\\
19336-0400&$b$& nd & nd &0.54 & 13.83 & 0.03,0.01& - & - & - & $0.04\pm 0.01$ \\
20462+3416& 0.15 & 0.03 & 0.02 & 2.41 & 1.59 & 0.35,0.17 & $5\pm 2$ & $8\pm 4$ & $0.09\pm 0.01$ & $1.52\pm 0.01$ \\
22023+5249& 4.80 & 1.41 & 0.67 & 4.36 & 5.90 & 0.29,0.15 & $3.41\pm 0.03$ & $7.1\pm 0.1$ & $0.82\pm 0.01$ & $0.74\pm 0.01$ \\  
\hline
\multicolumn{11}{l}{nd : not detected} \\
\multicolumn{11}{l}{{\em a} : blended with other line(s)} \\
\multicolumn{11}{l}{{\em b} : the emission forms a constant background over the field} \\
\end{tabular}
\end{table*}

The flux in each emission line was obtained by performing aperture
photometry on each spectral channel contributing to the line, using
elliptical apertures adjusted to encompass all of the emission. These
contributions were then summed to give the total line flux.  This
approach has the advantage that each channel image is inspected and
artefacts (due to the IFU for example) can be recognized and excluded.

Deconvolved emission line images are presented for some objects and
these were obtained using the {\sevensize LUCY} task within the
{\sevensize STSDAS} package in {\sevensize IRAF}.  In each case,
individual spectral channels contributing to an emission line were
deconvolved using the corresponding (i.e same wavelength) channel from
the PSF reference star and the resulting deconvolved channels combined
to form the deconvolved line image.

\section{Results and analysis}

H$_2$ emission is detected strongly in 5 out of 6 objects;
IRAS~18062+2410, IRAS~18379-1707, IRAS~19200+3457, IRAS~20462+3416 and
IRAS~22023+5249. The emission forms spatially extended nebulae
exhibiting a variety of structures and we detect a rich H$_2$ spectrum
including several faint lines originating from high vibrational
states.  In the case of one object, IRAS~19336-0400\footnote{Hereafter
  we use abbrieviated IRAS designations for these objects, such as
  I19336.}, we see only weak H$_2$ 1--0~S(1), Q(1) and Q(3) lines
which appear constant over the NIFS field-of view, with no obvious
structure. The spectra are shown in Fig. 1 with key line fluxes and
ratios given in Table~2.  A complete list of detected H$_2$ lines is
given in Appendix B.

The $K$-band H$_2$ lines, and in particular their flux relative to the
1--0~S(1) transition, provide information on how the molecule is
excited. The presence of higher vibrational lines, such as 8--6~O(3),
means that the B-type central stars in these objects are
photo-exciting H$_2$. UV photons ($\lambda\approx 100$~nm) can excite
H$_2$ via the Lyman and Werner bands, followed by decay to a
vibrationally-excited level of the electronic ground state. Subsequent
electric quadrupole decays through lower ro-vibrational states produce
the H$_2$ spectrum, including the higher vibrational transitions that
we observe. A purely UV-pumped spectrum will result in characteristic
values for the ratios of certain lines, such as
1--0~S(1)/2--1~S(1)$=1.8$ and 1--0~S(1)/3--2~S(3)$=5.7$ (e.g. Black \&
Dalgarno 1976; Black \& van Dishoeck 1987).

\begin{figure*}
\label{spectrum1}
\epsfxsize=19cm \epsfbox{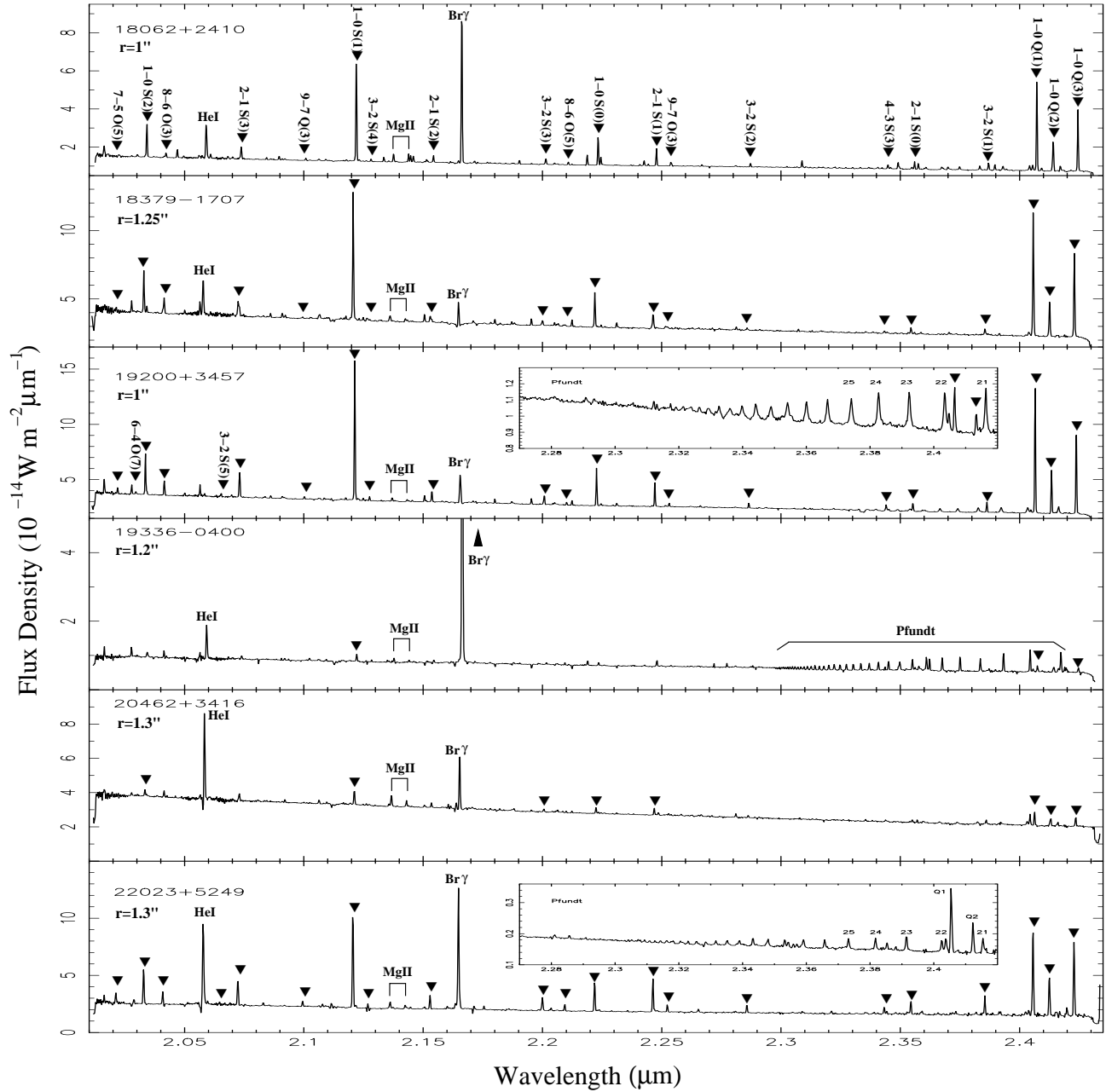}
\vspace*{-2cm}
\caption{{\sevensize NIFS} spectra between $2.01$ and
  $2.43$~$\umu$m. H$_2$ transitions are marked with inverted
    triangles as labelled in the top panel, with additional
  transitions marked where they occur. The {\sevensize He I}
  (2.058~$\umu$m) and Br$\gamma$ lines are indicated along with
    the Mg~{\sevensize II} doublet. The radius of the circular
    aperture used to extract each spectrum is indicated, in
    arcseconds. In IRAS 19200+3457, IRAS 19336-0400 and IRAS
    22023+5249 the Pfundt series is apparent. The inset boxes
    show spectra extracted from 0.5 arcsec radius apertures centred on
    the source.}
\end{figure*}

Our measured values for the 1--0~S(1)/2--1~S(1) and
1--0~S(1)/3--2~S(3) line ratios lie in the range 3--6 and 7--20
respectively (Table~2), so that at first sight they may not seem
consistent with UV-pumped excitation.  However, in dense gas subject
to intense UV radiation, excitation of the lower (${\rm v} \leq 1$)
vibrational levels of H$_2$ can be significantly affected by
collisional heating which acts to raise these ratios above their
radiative values (Sternberg \& Dalgarno 1989). The values we measure
can be obtained in photodissociation region (PDR) models with gas
densities greater than $\sim 10^{5}$~cm$^{-3}$ and UV intensities
$\sim 10^{4}$ times the interstellar value (Burton, Hollenbach \&
Tielens 1990). These values are likely to be encountered in our
objects. We estimate average densities in the ionized regions to be
$\sim 10^{4}$ to $\sim 10^{5}$~cm$^{-3}$ (Table~4) so that it does not
seem unreasonable to expect regions with density in excess of
$10^{5}$~cm$^{-3}$, especially if the gas is clumpy. The UV field at a
distance of $0.02$~pc (1~arcsec at 4~kpc) from a B1 post-AGB star
($T_{\rm eff}\approx 2\times 10^{4}$~K, $L\sim 10^{4}$~L$_\odot$) is
$\sim 10^{5}$ times the interstellar value, for 100~nm photons
(equation A7 in Sterberg \& Dalgarno 1989).

The lower vibrational levels of H$_2$ can also be excited thermally by
shocks in high velocity gas. The typical value for the
1--0~S(1)/2--1~S(1) ratio in molecular shocks is often quoted as
$\approx 10$, although a wide range of values is possible from 4
upwards (e.g. Shull \& Hollenbach 1978; Smith et al. 1995). However,
we see little evidence for shock structures in our objects, with the
possible exception of I22023 (Sec.~3.6). In general, the H$_2$
emission line spectra seem consistent with radiative excitation,
giving rise to the high vibrational level transitions observed,
combined with varying degrees of thermalisation of the ${\rm v}=1$
level due to heating by UV radiation from the B1 spectral type central
stars.

Br$\gamma$ emission is seen in all objects confirming that these stars
are beginning to photoionize their environments and create
H~{\sevensize II} regions.  In three cases (I19336, I20462 and I22023)
the Br$\gamma$ emission forms an extended nebula, giving an indication
of the geometry of the ionized region. I18062, I19336 and I22023 have
also been imaged at 8.4 GHz with the VLA (Cerrigone et al. 2008) and
in the latter two cases resolved structure is seen which is consistent
with that seen in Br$\gamma$.

We detect the $2.058$~$\umu$m He~{\sevensize I} line in all objects
except I19200. This line can be produced as part of the recombination
spectrum of He$^{+}$, indicating that the stars are becoming hot
enough to singly ionize He.  In I19336 we detect spatial structure in
the He~{\sevensize I}~2.058~$\umu$m emission and in two objects,
I18379 and I20462, this line is stronger than Br$\gamma$. The
He~{\sevensize I} 2.058/Br$\gamma$ ratio is discussed further in
Sec.~4.3.

In all objects apart from I19336 the continuum emission is not
obviously extended and instead is dominated by a central point source
(the FWHM of the PSF is listed in Table~1). In I19336 the extended
continuum is likely to be due to free-bound emission. This is in
contrast to earlier phase pre-PNe where the circumstellar material is
often revealed by dust-scattered star light. Dust scattering around
these more evolved hot post-AGB stars is not a major contributor to
their near-IR emission and therefore broadband imaging at these
wavelengths is not an effective tool for studying their circumstellar
material. Instead, outflows in this phase are seen in molecular and
atomic lines.

In Table~3 we give the heliocentric radial velocities obtained from
gaussian-fitting the line centres. In two cases, I18379 and I22023,
the H$_2$ emission shows velocity structure across the object. We now
discuss the results for individual objects in more detail.

\begin{table}
\caption{Heliocentric radial velocities in km~s$^{-1}$ ($\pm
  15$~km~s$^{-1}$) for Br$\gamma$ (2.16613~$\umu$m), He~{\sevensize I}
  (2.05885~$\umu$m; Benjamin, Skillman \& Smits 1999) and H$_2$
  $1-0$~S(1) (2.12183~$\umu$m; Black \& van Dishoeck 1987). $V_{\rm
    lit}$ gives the radial velocity reported in the literature. We
  detect velocity structure across the object in the $1-0$~S(1) line
  in I18379 and I22023 and so give the radial velocity range.
  References: 1. Mooney et al. (2002); 2. Smoker et al. (2004);
  3. Pereira \& Miranda (2007); 4. Garc\'{i}a-Lario et al. (1997);
  5. Sarkar et al. (2012). }
\begin{tabular}{lccccl}
IRAS ID & Br$\gamma$ & He~{\sevensize I} & 1--0~S(1) & $V_{\rm lit}$ & Ref\\
\hline 
18062+2410& 26 & 47 & 38 & $43\pm3$ & 1  \\
18379-1707& -156 & -132 & -188$\rightarrow$-109 & -133 & 2  \\
19200+3457& -64 & nd & -52 &  & \\
19336-0400& 55 & 93 & 75 & 203$\pm$34 & 3 \\
20462+3416& -92 & -57$^{a}$ & -67 & -75$\pm$6 & 4 \\
22023+5249& -170 & -140$^{a}$ & -185$\rightarrow$-146 & -148.3$\pm 0.6$ & 5  \\  
\hline
\multicolumn{6}{l}{nd : not detected} \\
\multicolumn{6}{l}{$a$ : this line has a blue-shifted absorption feature} \\
\end{tabular}
\end{table}

\subsection{IRAS~18062+2410}

This high galactic latitude ($b=+20$) post-AGB star (SAO 85766, HD
341617) appears to have evolved rapidly, from a spectral type of A5 in
the 1924 HDE catalogue to B1-1.5 based on a 1995 spectrum (Arkhipova
et al. 1999). The evolution may be even more dramatic, as
Parthasarathy et al. (2000a) point out that the spectral type was
still consistent with A5I in 1973 UV observations, but had become B1I
in their 1993 high resolution spectroscopy. This would correspond to
an incease in effective temperature from $\sim 8~500$ to $\sim
20~000$~K in 20 years.  Ryans et al. (2003) determine $T_{\rm
  eff}=20~750$~K and $\log g=2.35$, from high-resolution optical
spectroscopy and a non-LTE atmosphere model, placing the star close to
the $0.836$~M$_\odot$ post-AGB track of Bl\"{o}cker \& Sch\"{o}nberner
(1990). A high remnant mass would seem consistent with the rate of
spectral evolution, with the $0.836$~M$_\odot$ model
evolving from A5 to B1 in $\sim 50$~yr. Distance estimates range from
4.8 to 8.1~kpc (Arkhipova et al. 1999; Mooney et al. 2002).

The detection of low-excitation nebula emission lines such as [SII]
and [NII] (Arkhipova et al. 1999; Parthasarathy et al. 2000a; Arkhipova
et al. 2007) indicates that the object has just entered the PN
phase. This is consistent with the observed linear increase in the
8.4~GHz radio flux density from 1.46~mJy in 2001 (Umana et al. 2004)
to 2.7~mJy in 2009 (Cerrigone et al. 2009), which is interpreted as an
ionization front expanding at $\approx 120$~km~s$^{-1}$ through the
envelope (Cerrigone et al. 2011).  Their model predicts an ionized
region with outer radius $0.06$~arcsec assuming a distance of
$6.4$~kpc. The object was not resolved in 8.4~GHz VLA observations
with a beam size of 0.2 arcsec (Cerrigone et al. 2008). Based on the
linear increase in 8.4~GHz emission, these authors suggest that the
ionization began around 1991.

Optical spectroscopy shows carbon underabundance (Ryans et al. 2003;
Parthasarathy et al. 2000a) and observations with {\em ISO} and
subsequently {\em Spitzer} show strong amorphous silicate emission
features peaking at 10.8 and 17.6~$\umu$m, with no evidence for the
PAH features seen in C-rich or mixed chemistry sources (Gauba \&
Parthasarathy 2004; Cerrigone et al. 2009). Both of these authors model
the SED as a spherical shell of silicate dust using the {\sevensize
  DUSTY} code with an inner shell radius of $10^{16}$~cm (0.1~arcsec
at 6.4~kpc). This places the dust emission region just outside the
central ionized region. 

Cerrigone et al. (2009) estimate an envelope mass from the SED fit of
$3\times10^{-3}$~M$_\odot$, assuming a gas-to-dust ratio of
200. Although this will be an underestimate of the true envelope mass,
especially as there are no submm/mm SED constraints on the
contribution from cold dust, it is surprisingly low if the star has
evolved from an initial mass of $\sim 5$~M$_\odot$ to a post-AGB mass
of 0.836~M$_\odot$ and if the post-AGB evolution has taken $\sim 100$
years (Ryans et al. 2003).  Ryans et al. (2003) note that an error in
their estimation of $\log g$ of $0.3$ dex would place the star on a
significantly lower-mass post-AGB track, with a consequently longer
evolutionary timescale over which the envelope material may have
dissipated. However, the rapid increase in Br$\gamma$ and radio flux
lends support to a shorter evolutionary timescale.

Our {\sevensize NIFS} spectrum is shown in Fig.~1.  We see strong
Br$\gamma$ and He~{\sevensize I}~2$^{1}P-2^{1}S$ (2.058~$\umu$m)
lines, centred on the continuum peak and spatially unresolved,
associated with the central ionized region. The Br$\gamma$ flux has
increased by a factor of $5.3$ from the 1993 measurement of
Garc\'{i}a-Hern\'{a}ndez et al. (2002) to $4.35 \times
10^{-17}$~W~m$^{-2}$ in our 2007 NIFS data. KH05, measure an
intermediate flux of $2.48 \times 10^{-17}$~W~m$^{-2}$ in 1999, so
that the increase in Br$\gamma$ between 1993 and 2007 is
linear. Projecting this trend to earlier times suggests that
photoionization began around 1990. This is consistent with the linear
trend in the 8.4~GHz flux over the same period; as the ionization
front expands through the envelope, the free-free and recombination
emission should scale in the same way with the increasing volume of
ionized material. 

We have combined the continuum channels, avoiding prominent emission
lines, to produce an image of the continuum emission. However, we find
that there is no indication of an extended dust-scattered continuum in
I18062; this is again suprising if the star has lost several solar
masses of material on the AGB and then rapidly evolved through the
post-AGB stages. This is a common theme for the
hot post-AGB stars in this study; although we detect
extended molecular and/or atomic emission, there is little evidence
for any extended dust scattering region.

\begin{figure}
\label{18062-h2}
\epsfxsize=10cm \epsfbox[-48 0 474 687]{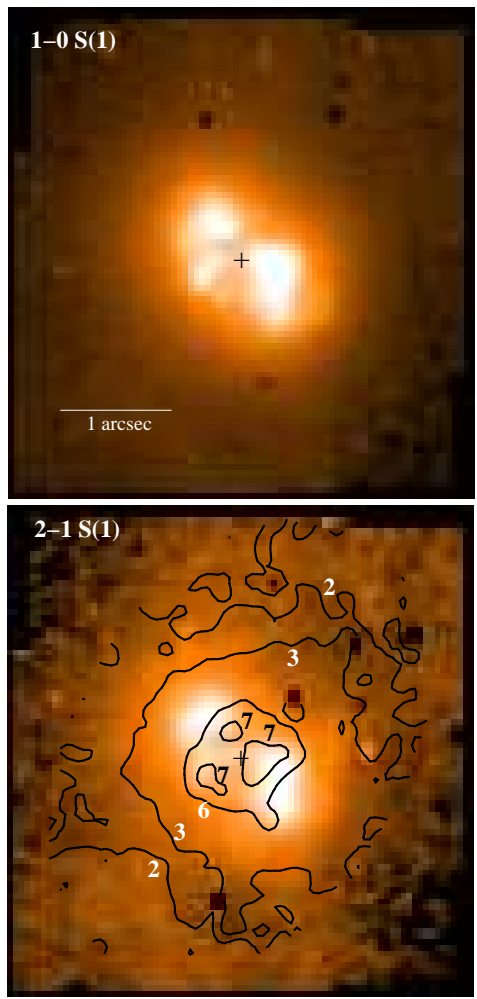}
\caption{Continuum-subtracted H$_2$ emission in IRAS 18062+2410
  displayed on a square root scale to show both bright and faint
  structure. The position of the continuum centroid is marked. Top:
  1--0~S(1) emission with white corresponding to $2.7 \times 10^{-17}$
  W~m$^{-2}$~arcsec$^{-2}$. Bottom: 2--1~S(1) emission with white
  corresponding to $4.5 \times 10^{-18}$
  W~m$^{-2}$~arcsec$^{-2}$. Contours of the 1--0~S(1)/2--1~S(1) line
  ratio are overlaid and labelled. The ratio varies from less than 2
  in outer regions to greater than 7 in central regions.  All images
  have N up and E left.}
\end{figure}

Fig.~1 shows that all v=1--0, 2--1 and 3--2 H$_2$ transitions
within the wavelength range are present. In addition we see higher
vibrational lines such as 4--3~S(3), 8--6~O(3) and 9--7~O(3).  We
measure a flux of $3.09 \times 10^{-17}$~W~m$^{-2}$ in the 1--0~S(1)
line, which, in contrast with Br$\gamma$, is the same as the 1993
measurement of Garc\'{i}a-Hern\'{a}ndez et al. (2002) and close to the
1999 value ($3.6 \times 10^{-17}$~W~m$^{-2}$) obtained by KH05. In
1993 the 1--0~S(1) line was 3.8 times stronger than the Br$\gamma$
whereas in 2007 Br$\gamma$ was 1.4 times stronger.

Fig.~2 (top) shows the continuum-subtracted 1--0~S(1) emission,
providing the first image of the extended nebula in this object.  The
centroid of the continuum and recombination line emission is marked
with a plus sign, scaled to have the same size as the central
radio-emitting region (radius 0.06 arcsec from Cerrigone et al. 2011).
On either side of this position we see peaks of H$_2$ emission, with
clumpy structure. The peaks are offset by $\approx 0.25$~arcsec from
the continuum centroid. Images of the higher vibrational level
transitions, such as 2--1~S(1), 3--2~S(3) and even 8--6~O(3), have a
very similar appearance so that the H$_2$ emission from these peaks is
at least partly due to UV pumped fluorescence. Faint extended emission
can be seen in the 2--1~S(1) line (Fig.~2 bottom), along PA
$318$\degr.  This morphology is similar to that of IRAS~19306+1407, a
B0/B1 post-AGB star, where arcs of clumpy H$_2$ emission are seen on
either side of the star and along a line perpendicular to a
larger-scale bipolar nebula (Lowe \& Gledhill 2006). A key difference
is that in the case of I19306, the arcs correspond to a dusty torus
seen in near-IR polarized light images (Lowe \& Gledhill 2007)
whereas, as noted above, we do not see any evidence for extended
dust-scattered light in I18062. We also obtained imaging polarimetry
for I18062 in 2006 using the UK Infrared Telescope and the same set-up
as for I19306, but failed to detect any intrinsic polarization in the
$J$- or $K$-bands.

The 1--0~S(1)/2--1~S(1) and 1--0~S(1)/3--2~S(1) line ratios are in the
range $5-9$ and $15-27$ respectively in the region of the peaks,
decreasing to $<2$ and $<5$, typical of a UV-pumped spectrum, in the
fainter outer regions (as measured from line ratio images.  Contours
of the 1--0~S(1)/2--1~S(1) line ratio are shown in Fig.~1.  These
ratios are consistent with the lower vibrational levels being
thermally excited to some degree in the region of the peaks, where the
gas density and UV intensity are expected to be higher, whereas in the
lower density outer regions of the nebula further from the UV source,
they retain their radiative values.

Davis et al. (2003) measured $5\pm5$ and $20\pm8$, respectively, for
1--0~S(1)/2--1~S(1) and 1--0S(1)/3--2~S(1) in 2001, using an
E-W-oriented $1.2$~arcsec-wide slit which would have included the
central peaks. However, the 1999 line ratios quoted by KH05 are 3.8
and 9.6, which are more comparable with our line ratios for the outer
faint material. This is due to their 2--1~S(1) and 3--2~S(1) line
fluxes being significantly higher than ours. Although KH05 used a
wider ($2.4$~arcsec) slit than Davis et al. (2003), their measurements
should still be dominated by the bright central emission. It is not
clear why these line ratios should have decreased in the 2 years
between the measurements of KH05 and Davis et al., and then remained
constant until our 2007 measurement.

We also note emission at 2.1375~$\umu$m, centred on the star and
spatially unresolved, which we believe to be a component of the
Mg~{\sevensize II} doublet (rest wavelengths 2.1375 and
2.1438~$\umu$m).  The longer wavelength component is blended with
other unidentified lines.  The Mg~{\sevensize II} doublet is common in
hot B-type stars with photoionized regions and is present in all our
spectra (Sec. 4.2).
 
\subsection{IRAS 18379-1707}
\begin{figure}
\label{18379-h2}
\epsfxsize=7cm \epsfbox{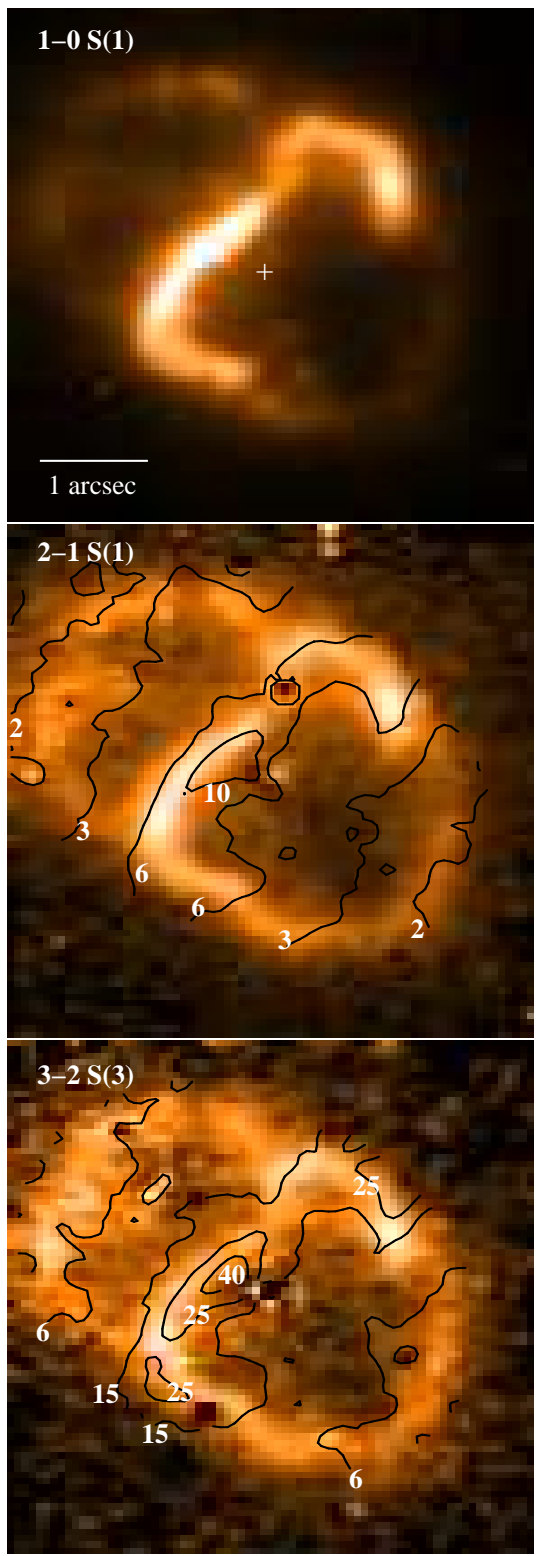}
\caption{Continuum-subtracted images of IRAS~18379-1707 in the
  1--0~S(1) (top), 2--1~S(1) (middle) and 3--2~S(3) (bottom) lines. 
  The greyscale peak (white) corresponds to
  $6.96\times10^{-17}$, $9.10\times 10^{-18}$ and $3.21\times
  10^{-18}$~W~m$^{-2}$~arcsec$^{-2}$ respectively. The position of the
  continuum centroid is marked with a plus symbol in the top image. Contours of
the 1--0~S(1)/2--1~S(1) and 1--0~S(1)/3--2~S(3) line ratios are shown in the 
middle and bottom panels, respectively.}
\end{figure}
I18379 (LSS 5112) is classified as a hot post-AGB star by
Parthasarathy, Vijpurkar \& Drilling (2000) with a spectral type of
B1IIIpe and {\it IRAS} colours typical of pre-PNe and PNe. Gauba \&
Parthasarathy (2004) model the double-peaked infrared SED as a
spherically-symmetric detached circumstellar envelope and determine an
inner angular radius of 0.64~arcsec, assuming silicate dust grains.
Spitzer observations by Cerrigone et al. (2009) show the dust to have
a mixed chemistry, with emission in the near-IR PAH bands as well as
the 10~$\umu$m silicate feature. These authors also present a
{\sevensize DUSTY} model of the continuum and SED, using a
silicate-only dust composition, but a hotter stellar temperature
($T_{\rm *}=$24~000~K) than Gauba \& Parthasarathy (2004) ($T_{\rm
  *}=$19~000~K). Interestingly the object is not detected in the radio
at 3.6~cm (Umana et al. 2004).

The spectrum (Fig.~1) shows Br$\gamma$ and He~{\sevensize I} emission,
which is centrally located and spatially unresolved, indicating a
still-compact ionized region. Br$\gamma$ is weaker than He~{\sevensize
  I} in our data ($0.67$ and $1.33\times 10^{-17}$~W~m$^{-2}$
respectively), although the Br$\gamma$ line sits on a depression in
the continuum which may lead to a slight underestimate of the flux;
KH05 quote $0.89\times 10^{-17}$~W~m$^{-2}$ for their 1999 Br$\gamma$
measurement.

\begin{figure}
\label{18379-vel}
\epsfxsize=7.5cm \epsfbox{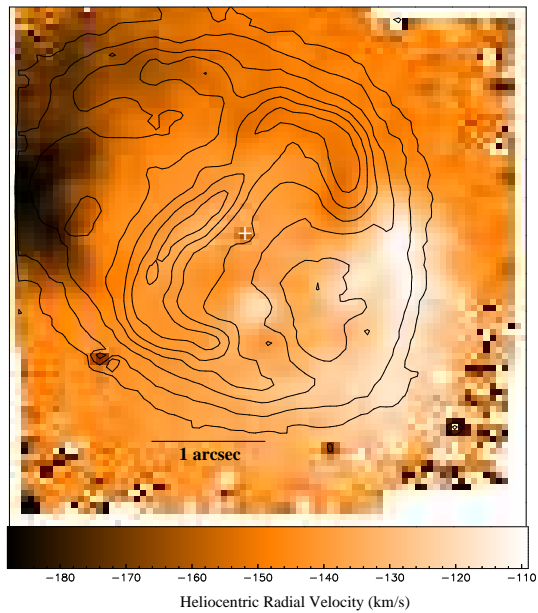}
\caption{A radial velocity image of the 1--0~S(1) emission in
  IRAS~18379-1707, obtained by gaussian-fitting the line centre,
  overlaid with contours of line intensity. The NE and SW rims of the
  nebula are relatively blue and red shifted, with a velocity
  separation of 80 km~s$^{-1}$. }
\end{figure}

KH05 detect H$_2$ emission with a E-W spatial extent of 3~arcsec,
including the 1--0~S(0) and S(1), 2--1~S(1) and 3--2~S(3) lines, and a
1--0~S(1)/2--1~S(1) line ratio of 3.1. They suggest a thermal
contribution to the H$_2$ excitation of 40 per cent. KH05 note an
unidentified line at 2.154~$\umu$m, which is not present in our
spectrum. However this is close to the 2--1~S(2) line which extends to
2.1535~$\umu$m (the H$_2$ lines are doppler broadened, see below). We
detect all the H$_2$ lines identified in I18062, with the addition of
7--5~O(5). We also note Mg~{\sevensize II} emission at 2.1362 and
2.1426~$\umu$m, centred on the star and spatially unresolved.

The detailed structure of the H$_2$ nebula can be seen in the NIFS
images shown in Fig.~3, each summed over the width of the respective
line. Hrivnak, Kelly \& Su (2004) show a {\em HST} narrowband image of
the 1--0~S(1) emission, with a ring-like structure which the authors
interpret as an equatorial disc. Our images show that the emission
takes the form of an oval shell of dimensions $3.6 \times
2.2$~arcsec oriented with long axis at 38\degr E of N, and a
bar-like feature across the minor axis of the shell which appears to
curve round at the ends and merge into the rim of the SW half of the
shell. In the 1--0~S(1) line the equatorial bar is much brighter than
the shell, but this contrast diminishes for the higher vibrational
transitions. 

The 1--0~S(1)/2--1~S(1) and 1--0~S(1)/3--2~S(3) line ratios are shown
as contours in Fig.~3 and have values of typically $6-10$ and $20-40$
respectively in the bar, but drop to 2 and $5$ at the NE and SW tips
of the shell. As in the case of I18062, this is consistent with high
densities ($>10^{4}$~cm$^{-3}$) and UV intensities ($>10^{2}$ times
interstellar) in the inner regions, which act to drive the line ratios
from their UV-pumped values towards thermal values (Sternberg et
al. 1989; Burton et al. 1990). In the outer shell the line ratios
retain their UV-pumped values as would be expected in a lower density
and less UV-intense environment further from the source.

We detect radial velocity structure in the H$_2$ emission across this
object spanning 80~km~s$^{-1}$ (Table~3). Fig.~4 shows a ``velocity
map'' obtained by gaussian-fitting the 1--0~S(1) line centre at each
spatial position and converting to heliocentric radial velocity. The
structure would be consistent with a bipolar outflow tilted so that
the NE/SW rims are approaching/receeding.

\subsection{IRAS 19200+3457}
\begin{figure}
\label{19200_h2}
\epsfxsize=7cm \epsfbox[-50 0 343 800]{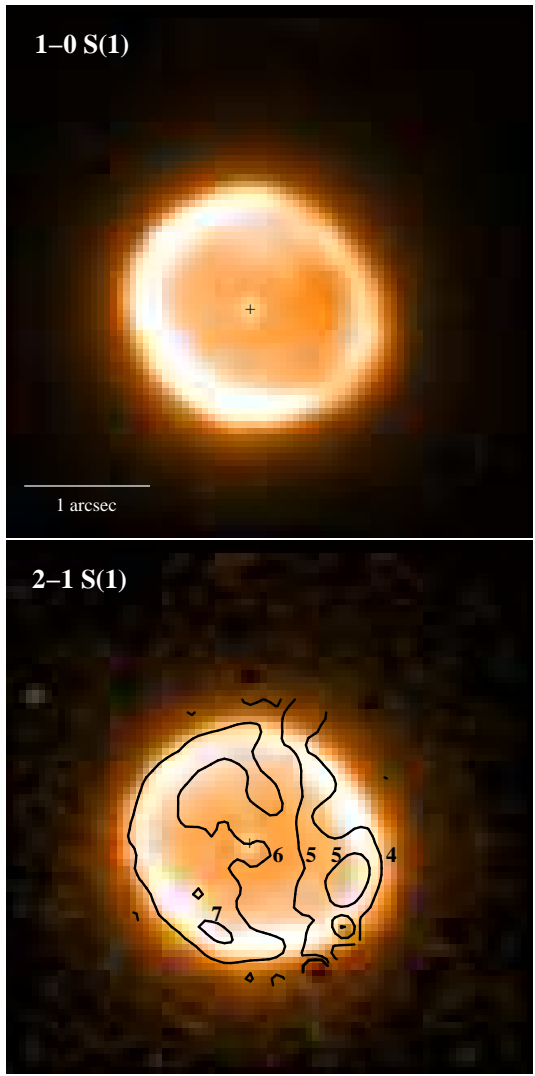}
\caption{Continuum-subtracted H$_2$ emission in IRAS 19200+3457. White
  represents $2.5 \times 10^{-17}$ and $4.1 \times
  10^{-18}$~W~m$^{-2}$~arcsec$^{-2}$ respectively for the 1--0~S(1) and
  2--1~S(1) images. Contours of the 1--0~S(1)/2--1~S(1) line ratio are overlaid
on the 2--1~S(1) image. The continuum centroid is marked.}
\end{figure}
This object is listed as a H$\alpha$ emission-line star (StHA 161) in
the catalogue of Stephenson (1986).  Optical and infrared photometry
(Gauba et al. 2003) shows a double-peaked SED, typical of a post-AGB
object with a dust envelope. Low-resolution spectroscopy (Arkhipova et
al. 2004) indicates a B-type star, with hydrogen lines in emission
which are variable and associated with the circumstellar
envelope. These authors note that the star shows irregular brightness
changes, similar to I18062. Emprechtinger et al. (2005) and Suarez et
al. (2006) also detect H$\alpha$ in emission, however Suarez et
al. (2006) give the object a Fe spectral classification.  {\em
  Spitzer} spectroscopy (Cerrigone et al. 2009) shows strong near-IR
emission features often attributed to PAHs as well as a broad
30~$\umu$m feature that may be due to MgS. The absence of silicate
features leads them to classify the object as C-rich. KH05 detect
H$_2$ emission with a line ratio of 1--0~S(1)/2--1~S(1)$=4.0$,
inferring a thermal origin for $50-60$ per cent of the
H$_2$ emission. They measure a Br$\gamma$ flux of $1.12 \times
10^{-17}$~W~m$^{-2}$.

We show the first resolved images of the circumstellar envelope in
Fig.~5; the H$_2$ emission takes the form of a roughly-circular ring
with an average radius of about 0.8~arcsec. The interior of the ring
contains emission, so that a likely configuration is a thin shell with
edge brightening. Alternatively, but less likely, the object may have
a bipolar structure oriented pole-on. We discuss the morphology
  of this object in more detail in Sec.~4.4.

The ring deviates from circularity, bulging out to the SW slightly.
The distribution of 1--0~S(1) and 2--1~S(1) emission around the ring
also differs; the W side of the ring is fainter in 1--0~S(1) than the
E side, whereas this is not the case for 2--1~S(1). This leads to a
1--0~S(1)/2--1~S(1) ratio of $6-7$ in the E side and $3-5$ in the W
side (see Fig.~5). The average value of $5.6$ is comparable to that of
KH05. The equivalent values for the 1--0~S(1)/3--2~S(3) ratios are
$12-18$, $8-12$ and $14.0$.

The nebular structure is similar in other lines of H$_2$, and all of
the stronger lines are detected (Fig.~1). We also detect Br$\gamma$
with a flux of $1.59 \times 10^{-17}$~W~m$^{-2}$ but do not detect the
2.058~$\umu$m line of He~{\sevensize I}. This is the only object in
our sample in which the He~{\sevensize I} line is not detected. The
Br$\gamma$ emission is not spatially resolved, and has the same
centroid and FWHM as the continuum. We detect the Mg~{\sevensize
  II} doublet lines at $2.1369$ and $2.1433$~$\umu$m.  A smaller
aperture centred on the star shows the hydrogen Pfundt series (inset
spectrum).

\subsection{IRAS 19336-0400}
\begin{figure*}
\label{19336}
\epsfxsize=18cm \epsfbox{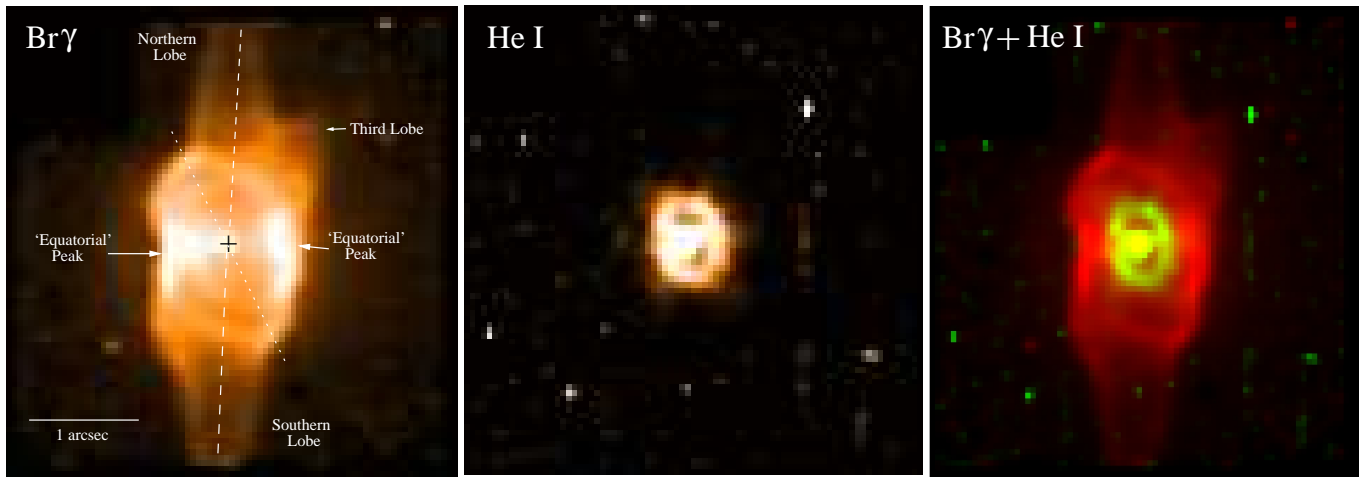}
\caption{IRAS 19336-0400. Left: Continuum-subtracted and deconvolved
  Br$\gamma$ image, displayed with a square-root scale to show both
  bright and faint structures. White corresponds to flux greater than
  $1\times 10^{-16}$~W~m$^{-2}$~arcsec$^{-2}$. Centre:
  Continuum-subtracted and deconvolved image of the $2.058$~$\umu$m
  He{\sevensize I} line, on a linear scale with white corresponding to
  flux greater than $5\times
  10^{-18}$~W~m$^{-2}$~arcsec$^{-2}$. Right: A composite image to
  illustrate the relative location of the Br$\gamma$ and He{\sevensize
    I} emission.}
\end{figure*}
A summary of some of the properties of this object from the literature
is given by Arkhipova et al. (2012); they present UBV photometry and
find that the star exhibits the irregular brightness fluctuations seen
in other B-type post-AGB objects, which they attribute to an unsteady
stellar wind. Optical spectroscopy shows Balmer lines in emission
as well as forbidden lines such as [N{\sevensize II}], [S{\sevensize
    II}], [O{\sevensize I}], suggesting a B1 supergiant and a very young
and low-excitation PN (Downes \& Keyes 1988; Van de Steene et
al. 1996; Parthasarathy et al. 2000b; Pereira \& Miranda 2007). The
object is detected in the radio with a flat spectrum between 1.4 and
8.4~GHz ($21-3.6$~cm) (Van de Steene \& Pottasch 1995; Umana et
al. 2004) indicating an optically thin nebula at these frequencies and
being consistent with the earliest stages of the ionization-bounded
phase (Umana et al. 2004). Further radio monitoring by Cerrigone et al.
(2011) shows that the radio flux has not varied significantly between
2001 and 2009. Observations with {\em Spitzer} show a mixed chemistry
with both PAH-like and silicate features evident, with crystalline
silicate features also present (Cerrigone et al. 2009).

KH05 report that this is the only object in their sample where the
Br$\gamma$ line is stronger than the 1--0~S(1) H$_2$ line with fluxes
of $10.7 \times 10^{-17}$ and $1.01 \times 10^{-17}$~W~m$^{-2}$
respectively. We detect weak 1--0~S(1), Q(1) and Q(3) lines, but the
emission appears constant across the NIFS field-of-view, with no
discernible structure. This is consistent with KH05's observation that
``the H$_2$~1--0~S(1) line extends for 11~arcsec along the slit, with
a fairly constant brightness distribution''. As we have imaged only a
small part of the H$_2$ emission, we do not give a flux measurement.
Our Br$\gamma$ flux is $13.8 \times 10^{-17}$~W~m$^{-2}$. In Fig.~6
(left) we show the Br$\gamma$ image, deconvolved with the standard
star and with a square-root scaling to display both the bright and
faint emission. The object is clearly bipolar, with two narrow,
tapering edge-brightened lobes, extending beyond the edge of our field
($> \pm 2$~arcsec from the star) at a PA of 358\degr~E of N (shown as
a dashed line in the Figure). There is evidence for a third, smaller
lobe to the NW of the star. The brighter part of the nebula is bounded
by a rim of emission, which appears stretched along a PA of 26\degr,
shown by the dotted line. This line joins two brighter spots of
emission on the rim. These various structures and axes give the object
a multi-polar appearance. The brightest emission lies along a line
perpendicular to the main lobes, most likely corresponding to an
enhancement in the gas density in the plane perpendicular to the
bipolar axis. These bright ``equatorial peaks'' can be clearly seen in
the $8.4$~GHz image of Cerrigone et al. (2008). We have also
deconvolved the He~{\sevensize I} image using the standard star, to
reveal a ring of emission (Fig.~6 middle) with extent $0.4
\times0.3$~arcsec and PA 10\degr~E of N. The He~{\sevensize I} ring
sits within the cavity of the Br$\gamma$ nebula, as shown in the
righthand panel.

The continuum image of the object is dominated by the bright central
star although there is faint emission, with structure similar to that
of the Br$\gamma$ image, so this is likely due to free-free and
free-bound continuum rather than scattering from dust.

We again see the Mg~{\sevensize II} emission lines, at 2.1378 and
2.1442~$\umu$m. The separation is $6.4\times 10^{-3}$~$\umu$m and the
flux ratio approximately 2:1, as for the other objects. Although the
lines are faint, the emission is strongest in the ``equatorial
peaks''.  The Pfundt series is visible to the 21--5 line at
$2.4173$~$\umu$m; images of the Pfundt emission have a similar
structure to that of the Br$\gamma$ emission, as expected.

\subsection{IRAS 20462+3416}

\begin{figure}
\label{20462_BrG}
\epsfxsize=7cm \epsfbox{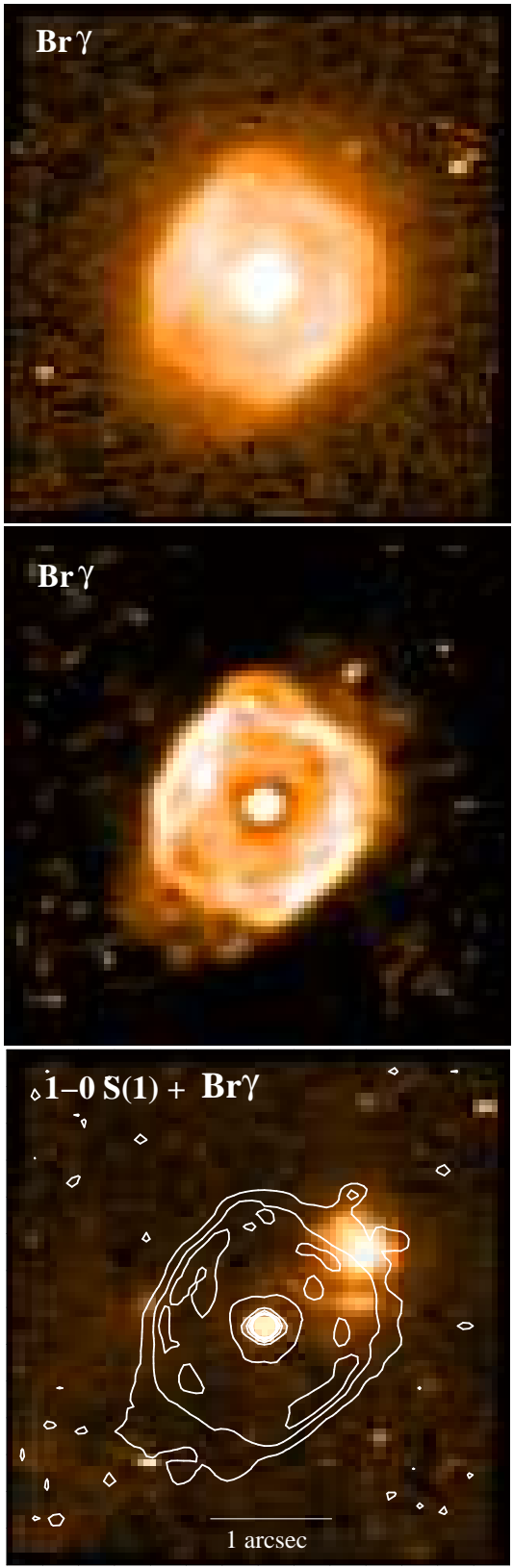}
\caption{IRAS 20462+3416: An image of the continuum-subtracted
  Br$\gamma$ emission (top) and its deconvolution using a
  standard star (middle). The
  two images are displayed on the same square-root-compressed scale
  with white corresponding to $6.54 \times
  10^{-18}$~W~m$^{-2}$~arcsec$^{-2}$. Bottom:
  1--0~S(1) H$_2$ emmision, with white corresponding to $3.65 \times
  10^{-18}$~W~m$^{-2}$~arcsec$^{-2}$ overlaid with selected Br$\gamma$
  contours.}
\end{figure}

I20462 was identified as a hot post-AGB star at a distance of
$2.9-4.6$~kpc (Parthasarathy 1993), having previously been classified
as a population I B1.5 supergiant at large distance (Turner \&
Drilling 1984). Photometric and spectroscopic variations are thought
to be due to stellar pulsations and ongoing mass loss
(Garc\'{i}a-Lario et al. 1997; Arkhipova et al. 2001). The mid-IR
spectra show PAH as well as crystalline silicate features (Cerrigone
et al. 2009). The object is detected at radio frequencies between 1.4
and 22.2~GHz with a flat spectrum (spectral index -0.02) apart from
the 22.4~GHz measurement which shows an excess possibly due to cold
dust (Cerrigone et al. 2008, 2009).

We detect strong Br$\gamma$ and He~{\sevensize I} emission from this
object, with fluxes of $1.59$ and $2.41 \times 10^{-17}$~W~m$^{-2}$
respectively. The Br$\gamma$ is clearly extended, forming a boxy
nebula with dimensions $\approx 2.0\times 1.5$~arcsec oriented along a
PA of 49\degr~(Fig.~7 top). We have deconvolved the image using the
standard star, to highlight the edge-brightened structure of the
nebula and also reveal faint emission extending to the NW and SE
(Fig.~7 middle). The object is also extended in the $V$ and $I$ band
{\em HST} imaging of Ueta et al. (2000), forming a faint elliptical
nebula of $\approx 4\times 3$~arcsec with a similar orientation. The
Br$\gamma$ nebula appears to sit inside the optical nebula, which has
a limb-brightened shell appearance in the {\em HST} images.

The 1--0~S(1) image is shown in Fig.~7 (bottom).  There appears to be
faint H$_2$ emission covering our $4 \times 4$~arcsec field, which is
consistent with KH05's observation that the emission extends over
12~arcsec along their $2.4$~arcsec-wide EW-oriented slit. There is
also a concentration of H$_2$ emission $\approx 0.75$~arcsec to the NW
of the star with an arc-like feature below it, located along the major
axis of the extended Br$\gamma$ nebula (shown contoured in the
Figure). The faint emission covering the field is blue-shifted by one
spectral channel ($2.13\times 10^{-4}~\umu$m) relative to the other
H$_2$ features, and may represent the front surface of a molecular
envelope, surrounding the dust shell and ionized region.

All 1--0 transitions in the wavelength range are detected, along with
2--1~S(1) and 3--2~S(3). The flux measured in the 1--0~S(1) line
is $1.49 \times 10^{-18}$~W~m$^{-2}$, and the 1--0~S(1)/2--1~S(1) line
ratio is $5.4$.  The He~{\sevensize I} emission is spatially
unresolved in our observations and displays a clear blue-shifted P
Cygni absorption feature, with a velocity separation of
$-93$~km~s$^{-1}$ from the emission peak, or $-155$~km~s$^{-1}$ from
the peak to the blue edge of the absorption. P Cygni profiles are also
seen in optical H and He lines and indicate ongoing mass loss from the
star (Garc\'{i}a-Lario et al. 1997; Arkhipova et al. 2001).

KH05 report ``additional lines'' at $2.138$ and $2.287$~$\umu$m; the
$2.287$~$\umu$m line may be H$_2$ $3-2$~S(2), which we do not
detect. As in all our targets, we also detect the Mg~{\sevensize II}
doublet in this object, at $2.1368$ and $2.1430$~$\umu$m with an
intensity ratio of approximately 2:1 (Table 2).  Both these lines
appear spatially extended beyond the central peak, with faint emission
extending over the region covered by the Br$\gamma$ nebula.

\subsection{IRAS 22023+5449}
\begin{figure*}
\epsfxsize=18cm \epsfbox{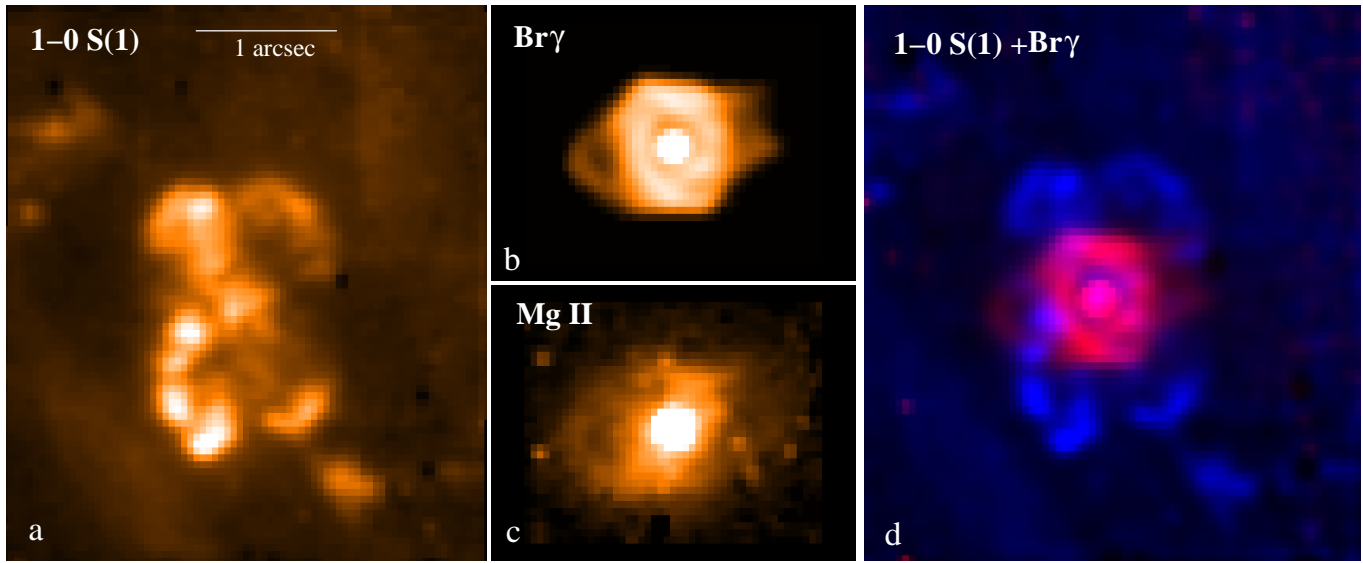}
\caption{Continuum-subtracted line images for IRAS~22023+5249: 
(a) 1--0~S(1) H$_2$ line, scaled
  between 0 and $2\times10^{-17}$~W~m$^{-2}$~arcsec$^{-2}$; 
(b) Lucy-Richardson deconvolution of the Br$\gamma$
  line, displayed with a logarithmic stretch to highlight faint
  structure, between 0 and
  $1.2\times10^{-16}$~W~m$^{-2}$~arcsec$^{-2}$;(c) a combined
  image of the Mg~{\sevensize II} lines at 2.1362 and 2.1424~$\umu$m, which show similar
  extension and structure to the Br$\gamma$ line, scaled between 0 and
  $7\times10^{-19}$~W~m$^{-2}$~arcsec$^{-2}$; (d) A combination of
  the H$_2$ image in blue and Br$\gamma$ image in red, showing the
  relative location and extent of the emission.}
\end{figure*}
Sarkar et al. (2012) classify this object as a hot B0-B1 post-AGB
star, with $T_{\rm eff}=24~000$~K, $\log g=3.0$, using a
high-resolution ($R\sim 50~000$) optical spectrum. They note P Cygni
profiles in the Balmer and He lines, suggesting on-going mass
loss. Multi-frequency radio observations give a spectral index $\sim
-0.1$ in agreement with an optically thin shell model (Cerrigone et
al. 2011). KH05 detect a number of H$_2$ emission lines with spatial
extent $\sim 7$~arcsec including the 9--7~O(3)
transition. They also note that the Br$\gamma$ line has an extent of
2--3~arcsec.

The H$_2$ spectrum (Fig.~1) is particularly rich, with all lines
listed by Black \& van Dishoeck (1987) in our wavelength range
detected, including vibrational states up to v=10. We see strong
Br$\gamma$ and Pfundt series lines (inset spectrum) and the weaker
Mg~{\sevensize II} doublet. The He~{\sevensize I} line at
2.058~$\umu$m is strong and has a P Cygni absorption profile,
blue shifted from the line peak by $-124$~km~s$^{-1}$.

In Fig.~8(a) we show the continuum-subtracted image of the 1--0~S(1)
line. As well as faint diffuse emission extending over much of the
NIFS field, we see a number of bright distinct knots and curved
features, reminiscent of bow shocks, arranged around the stellar
centroid. Two fainter knots are offset $1.5$~arcsec from the star to
the NE and SW. The Br$\gamma$ emission is also spatially extended (as
noted by KH05) and a deconvolution using the standard star is shown in
Fig.~8(b). The Br$\gamma$ forms a 0.6 arcsec-diameter ring-like
structure around the star, with a fainter elliptical shell or bubble
extending along PA 100\degr, which appears open to the NW. There is a
central unresolved peak of Br$\gamma$ emission.  The Br$\gamma$ image
(red) and 1--0~S(1) image (blue) are shown together in Fig.~8(d) to
illustrate the relative location and extent of the photoionized region
and molecular emission.  We also see extended emission from the
Mg~{\sevensize II} lines at 2.1362 and 2.1424~$\umu$m, with a
structure similar to that of the Br$\gamma$ line, showing an extended
elliptical shell or bubble but without the ring structure seen in
Br$\gamma$; we show the summed images of these two lines in
Fig.~8(c). The He~{\sevensize I} emission is centrally located and not
resolved.

We would expect the free-free emission seen at radio wavelengths to
show similar structure to the H recombination lines, and indeed most
of the 8.4~GHz emission is located in a central 1 arcsec-diameter
region (Cerrigone et al. 2008). However, the 8.4~GHz image also shows
two protrusions of emission extending out to $\approx 1$~arcsec from
the star along PA$\approx 135$\degr, which do not appear to correspond
to any of the recombination line structures seen in our data. It is
possible that these extensions represent structures distinct from the
photoionized region, and instead may be related to jets, as suggested
by Cerrigone et al. (2008).

In constrast to the v=1--0 H$_2$ transitions, which are strong in the
central knots and arcs, higher vibrational transitions appear more
prominent in the diffuse nebulosity at larger angular offset from the
centre. This is shown in Fig.~9 for the 2--1~S(1) and 3--2~S(3) lines,
where in the latter case the knot structure is almost absent. This
leads to variation in emission line ratios across the object: the
1--0~S(1)/2--1~S(1) ratio is $> 8$ in the knots, and $\approx 2$ in
the diffuse nebulosity. The 1--0~S(1)/3--2~S(3) ratio varies from $>
20$ in the knots, and $\approx 4$ in the diffuse nebulosity. As
discussed for I18062 and I18379, this is consistent with a degree of
thermalization of the lower vibrational level populations in the
denser region of the knots, whereas the line ratios in the diffuse
regions have their UV-pumped values, typical of a low-density
gas. However, if jets are present in this object then heating of the
gas by shocks may be important. As mentioned earlier, some of the
knots resemble bow shocks.

We see radial velocity structure across the object in the H$_2$
lines (Fig. 10), with a split between blue-shifted emission to the 
W of the star and red-shifted emission to the E.

\section{Discussion}
\subsection{The photoionized region}
Br$\gamma$ emission is detected in all 6 objects, indicating that
photoionization of the circumstellar environment has begun. In I18062,
I18379 and I19200 the photoionized region is still too compact to
spatially resolve in our observations whereas for I19336, I20462 and
I22023 we see a Br$\gamma$ nebula. We can use the angular extent of
the ionized region and the Br$\gamma$ flux to estimate the hydrogen
number density $n_{\rm H}$ and the mass of ionized material $M_{\rm
  i}$, assumng a constant density spherically symmetric ionization
region (see Table 4 and Appendix A). 

For the 3 objects with resolved photoionized regions we find $n_{\rm
  H}\sim 10^{4}$~cm$^{-3}$ whereas the unresolved objects have higher
density, $n_{\rm H}\sim 10^{5}$~cm$^{-3}$ with I18062 being the most
dense, which would be consistent with it being in the early stages of
photoionization.  We have used a radius of 0.06 arcsec for the ionized
region in I18062, consistent with the Cerrigone et al. (2011) radio
model. They find that the ionized mass increased from $1.5\times
10^{-4}$ to $3.3\times 10^{-4}$~M$_{\odot}$ between 2003 and 2009,
which agrees well with our estimate, considering the approximate
nature of our calculation. For I18379 and I19200 we can only place
upper limits on the radius of the ionized region and hence lower and
upper limits, respectively, on the density and ionized mass.
\begin{table}
\caption{Hydrogen densities ($n_{\rm H}$) and ionized masses ($M_{\rm i}$) for 
our objects, assuming the given distances from the literature. See Section 
4.1 and Appendix A for details.}
\begin{tabular}{lcccc}
IRAS ID & $d$    & $\phi$  & $n_{\rm H}$ & $M_{\rm i}$ \\
        & kpc    & arcsec  & cm$^{-3}$  & $\times 10^{-4}$M$_\odot$ \\
\hline 
18062+2410 & 6.4$^{a}$ & 0.06$^{a}$ &$10^{5.6}$   & 2.0  \\
18379-1707 & 3.7$^{b}$ & $<0.2$     &$>10^{4.5}$  & $<1.2$ \\
19200+3457 & 3.8$^{c}$ & $<0.2$    &$>10^{4.6}$ & $<1.9$   \\
19336-0400 & 4.0$^{d}$ & 1.5       &$10^{3.8}$ & 140   \\
20462+3416 & 3.8$^{e}$ & 0.6       &$10^{3.9}$ & 10.6 \\
22023+5249 & 3.9$^{b}$ & 0.7       &$10^{4.1}$ & 27.3  \\  
\hline
\multicolumn{5}{l}{$^{a}$Cerrigone et al. (2011); $^{b}$Gauba \& Parthasarathy (2004);}\\
\multicolumn{5}{l}{$^{c}$Gauba, Parthasarathy et al. (2003); $^{d}$Cerrigone et al. (2008);}\\
\multicolumn{5}{l}{$^{e}$Parthasarathy (1993)}\\ 
\end{tabular}
\end{table}

Ionized masses are $\sim 10^{-4}$~M$_\odot$ and $\sim
10^{-3}$~M$_\odot$ for objects with unresolved and resolved ionized
regions respectively, with the exception of the most extended object
I19336. Bearing in mind the uncertain distances, this is consistent
with the unresolved objects being at a relatively early stage in
ionizing their envelopes. The ionized mass in I19336 is of the same
order of magnitude as the envelope mass (Cerrigone et al. 2009), so
that the ionization front has propagated through the nebula and most
of the molecular material has been dissociated (as indicated by the
very weak H$_2$ emission in this object).

It is, however, evident from the structure in these emission line
nebulae that the ionized regions are not spherically symmetric and the
assumption of constant density is not realistic, especially once the
ionization front encounters material shaped during the pre-PN
phase. The values in Table~4 are intended as order of magnitude
estimates.
\begin{figure}
\epsfxsize=8.5cm \epsfbox{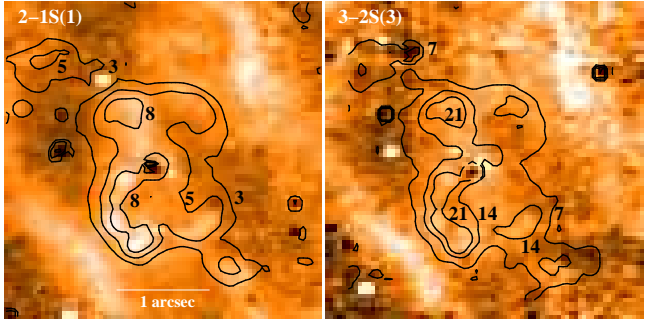}
\caption{Images of the 2--1~S(1) (left) and 3--2~S(3) (right) lines in
  IRAS 22023+5249.  White corresponds to $2\times10^{-18}$ and
  $2\times 10^{-19}$~W~m$^{-2}$~arcsec$^{-2}$ respectively. Contours
  of the 1--0~S(1)/2--1~S(1) and 1--0~S(1)/3--2~S(3) line ratios are
  superimposed on the 2--1~S(1) and 3--2~S(3) images, respectively.}
\end{figure}
\begin{figure}
\epsfxsize=7cm \epsfbox{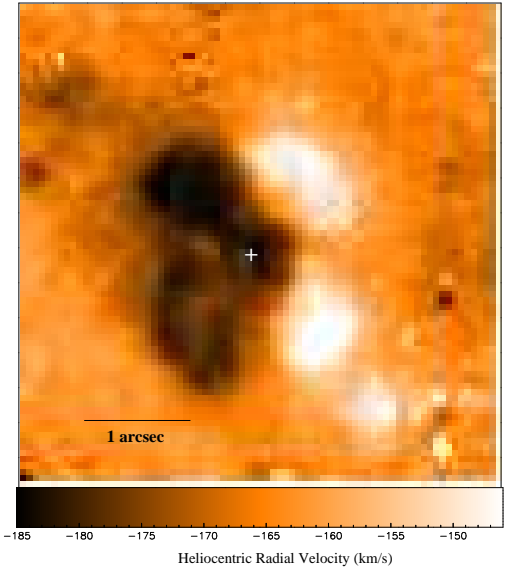}
\caption{A radial velocity image of the 1--0~S(1) emission in
  IRAS~22023+5249, obtained by gaussian-fitting the line centre. The E
  and W parts of the nebula are relatively blue and red shifted, with
  a velocity separation of up to 40 km~s$^{-1}$.}
\end{figure}

\subsection{Mg~{\sevensize II} lines}
The Mg~{\sevensize II} doublet, resulting from the
$5~p^{2}P_{3/2}\rightarrow 5~s^{2}S_{1/2}$ ($2.1375~\mu$m) and
$5~p^{2}P_{1/2}\rightarrow 5~s^{2}S_{1/2}$ ($2.1438~\mu$m)
transitions, is seen in each of our targets (see Table 2). The
$5~p^{2}P_{3/2}$ and $5~p^{2}P_{1/2}$ levels can be populated from the
$3~s^{2}S_{1/2}$ Mg$^{+}$ ground state by line overlap with Ly$\beta$
(Bowen 1947) with a velocity difference of 73 and 116~km~s$^{-1}$,
respectively (e.g. Simon \& Cassar 1984). The Mg~{\sevensize II}
doublet has been noted in the PN Hubble~12 (Hb~12) (Luhman \& Rieke 1996)
where the assumption of a Ly$\beta$ line wide enough to fluoresce both
levels leads to an expected Mg~{\sevensize II} line ratio of 2 to 1,
which is consistent with our observations.

The Mg~{\sevensize II} emission in our targets is very strong with
respect to Br$\gamma$, with a ratio greater than $0.05$ in all cases
apart from I19336, the object with the most extended region of ionized
emission, with Mg~{\sevensize II}/Br$\gamma=0.003$. Luhman \& Rieke
(1996) measure Mg~{\sevensize II}/Br$\gamma=0.006$ towards the centre
of Hb 12. Where the Br$\gamma$ emission is extended (I19336,
I20462, I22023), the Mg~{\sevensize II} emission shows the same
extended structure. The flourescence process appears to be very
efficient, suggesting that Ly$\beta$ is optically thick in these very
early PNe and that there are velocity gradients in the ionized
region allowing the line overlap to occur. 

\subsection{He~I 2.058~$\umu$m/Br$\gamma$ ratios}
Photons with energies greater than 24.6~eV ($\lambda < 0.0504~\umu$m)
are capable of ionizing He as well as H, and as the number of these
photons increases with increasing $T_{\rm eff}$ of the central star, a
volume of singly ionized He will develop at the centre of the H$^{+}$
region. Recombination of He$^{+}$ can populate the 2$^{1}P$ state with
the 2$^{1}P\rightarrow$2$^{1}S$ transition resulting in emission of a
He~{\sevensize I} $2.058$~$\umu$m photon (we use a rest wavelength of
$2.05885$~$\umu$m calculated from the energy levels quoted in Benjamin
et al. 1999).  The He$^{+}$ region is expected to extend throughout
the H$^{+}$ region for stars with $T_{\rm eff}$ greater than about
40~000~K (e.g. Shields 1993), corresponding to a spectral type of
$\sim$ O5I (Martins, Schaerer \& Hillier 2005).  For our cooler B1
stars, the radius of any He$^{+}$ zone should be considerably smaller;
if we assume that the ratio of the ionizing fluxes with $h\nu >
24.6$~eV and $h\nu > 13.6$~eV is $\sim 0.001$ for a B1I star (see
fig.~17 of Martins et al. 2005), a helium fraction $n_{\rm He}/n_{\rm
  H}\sim 0.1$, and spherical symmetry then, using equations 15.35 and
15.36 of Draine (2010), the radius of the He$^{+}$ zone will be $\sim
0.2$ that of the H$^{+}$ zone. This fits with our observations of
I19336, which has the most extended Br$\gamma$ nebula with an extent
of $\approx 4\times 2$~arcsec and a resolved He~{\sevensize I} extent
of $\approx 0.9\times 0.6$~arcsec. In the case of I20462 and I22023
which have $\phi$ equal to 0.6 and 0.7 arcsec respectively (Table~4),
the He~{\sevensize I} emission is still not spatially resolved in our
data.
 
We detect the He~{\sevensize I} line in all targets apart from I19200,
indicating that He$^{+}$ is present in these objects. The strength of
this line relative to hydrogen Br$\gamma$ will depend on the relative
volumes of recombining helium and hydrogen, and hence on the effective
temperature of the exciting star. However there are a number of issues
that complicate the interpretation of the He~{\sevensize
  I}~2.058~$\umu$m/Br$\gamma$ ratio. The 2$^{1}P$ level can also decay
to 1$^{1}S$ resulting in a $0.0584~\umu$m photon. Treffers et
al. (1976) point out that this line must be optically thick to avoid
depopulating the 2$^{1}P$ state; the $0.0584~\umu$m photons then
scatter within the nebula and are converted into $2.058$~$\umu$m
photons by a process of resonant fluorescence, substantially enhancing
the $2.058$~$\umu$m emission. The efficiency of this process is
reduced, however, if the short wavelength photons are intercepted by
dust or neutral hydrogen within the H~{\sevensize II} region (Shields
1993). Also, for gas densities greater than $\sim 10^{3}$~cm$^{-3}$,
collisional transfer from the metastable triplet 2$^{3}S$ state
becomes the principal route to populate 2$^{1}P$, so that the
$2.058$~$\umu$m emission scales with density as well as with $T_{\rm
  eff}$.

Depoy \& Shields (1994) calculate the He~{\sevensize
  I}~2.058~$\umu$m/Br$\gamma$ ratio for a grid of ($T_{\rm eff}$,
$n_{\rm H}$) values using the {\sevensize CLOUDY} photoionization code
and parameter values applicable to planetary nebulae. For a given
density, their He~{\sevensize I}~2.058~$\umu$m/Br$\gamma$ ratio
increases with $T_{\rm eff}$, peaking around $T_{\rm eff}=40~000$~K,
before declining again as He$^{+}$ is ionized to He$^{++}$ at high
temperatures.  For a given $T_{\rm eff}$, the He~{\sevensize
  I}~2.058~$\umu$m/Br$\gamma$ scales with $n_{\rm H}$ due to
collisional population of 2$^{1}P$.  For $T_{\rm eff}=25~000$~K (the
minimum temperature in their plot) we find that He~{\sevensize
  I}~2.058~$\umu$m/Br$\gamma > 1.0$ and $ > 2.0$ require $n_{\rm H} >
10^{4.1}$~cm$^{-3}$ and $ > 10^{4.5}$~cm$^{-3}$ respectively. The B1
stars in our sample will be cooler, $T_{\rm eff}\approx 20~000$~K, and
so $n_{\rm H}$ should be higher for the same line ratios. This is
consistent with our average density estimates (Table 4) where $n_{\rm
  H}$ is in the range $10^{3.8}\rightarrow 10^{5.6}$.

\subsection{Dust and morphology}

The SEDs of our NIFS targets are double-peaked, with a dust emission
peak in the mid-IR, between 20 and 30~$\umu$m (Cerrigone et al. 2009).
They belong to the Type IVb SED class of van de Veen, Habing \&
Geballe (1989) produced by post-AGB stars surrounded by optically thin
dust envelopes. At optical wavelengths, these objects often appear as
bright stars surrounded by faint reflection nebulosity, some with
multiple symmetry axes, and were termed SOLE (Star-Obvious
Low-level-Elongated) objects by Ueta et al. (2000); in mid-IR images
they may appear ``toroidal'' with two peaks of emission, one on either
side of the star (e.g Meixner et al. 1999).

Only one of our targets, I20462, has previous optical imaging; it
appears in $V$ and $I$ filter WFPC2 images as a faint elliptical
reflection nebulosity surrounding the bright star (Ueta et
al. 2000). If we assume that the dust-scattered intensity of the
nebulosity varies as a power law with wavelength, so that $I_{\lambda}
\propto \lambda ^{-p}$, then the values of Ueta et al. for the
specific intensity in the nebulosity in the $V$ and $I$ filters (their
Table 1) suggest that $p \approx 2.3$. Assuming that this wavelength
dependency continues into the IR then we would expect roughly $5\times
10^{-20}$~W~m$^{-2}$~arcsec$^{-2}$ due to scattered continuum per
spectral channel at $2.2~\umu$m. This is close to the noise level in
our NIFS data, and equivalent to about 10 per cent of the surface
brightness in Br$\gamma$ in the faint regions of the nebulosity
(Fig.~7). We have not detected any extended continuum above this level
outside the region of the Br$\gamma$ nebula.  Imaging polarimetry in
the $J$ band (Gledhill 2005) shows a faint reflection nebula of radius
$\sim 2$~arcsec around the star, with some evidence for a
concentration of scattering material along the minor axis of the
Br$\gamma$ and optical nebulae. I20462 is the only B-type SOLE object
in the compilation of Ueta et al. (2000), other targets being
proto-PNe with cooler F-G spectral type central stars. These younger
objects are clearly more dusty, with brighter reflection nebulosites,
also detected in the near-IR (Gledhill et al. 2001; Gledhill 2005).

None of the hot post-AGB objects in our NIFS sample show evidence for
extended dust envelopes, so they are similar in this respect to
I20462, and would likely appear at optical wavelengths as faint
reflection nebulosities around bright central stars. This similarity
is underscored by their location in the two-colour $J-K$
vs. $K-[25]$\footnote{$K-[25]$ compares the $K$ band flux to the {\em
    IRAS} $25~\umu$m flux, defined as $[25]=-2.5\log10 [F_{\nu}/6.73]$
  in Ueta et al. (2000).}  diagram, where they occupy similar
locations to I20462 (Fig.~11).  With the exception of I19200, the hot
post-AGB stars lie to the right of the pre-PNe (F to G spectral type)
SOLE objects, in a region also occupied by ionized PNe such as Hb~12.

\begin{figure}
\epsfxsize=8.5cm \epsfbox[0 110 543 681]{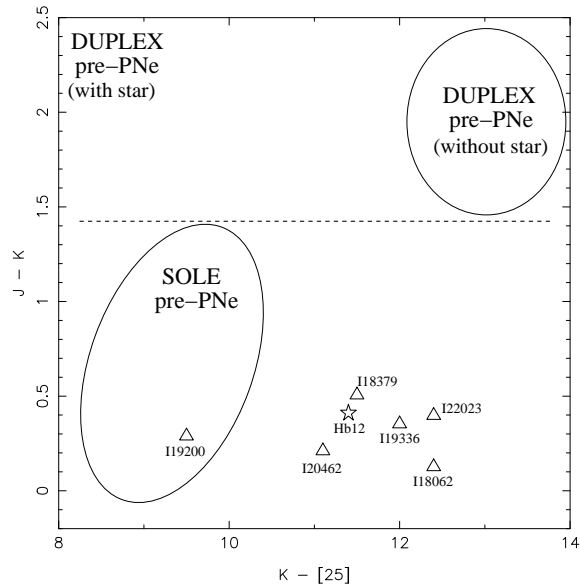}
\caption{$J-K$ versus $K-[25]$ two-colour plot showing the location of
  our sources (triangles) relative to the regions occupied by SOLE and
  DUPLEX pre-PNe (Ueta et al. 2000). The location of the PN Hb~12 is
  also marked.}
\end{figure}

Spectroscopy with {\em Spitzer} shows that I19200 is the only C-rich
object in our sample, I18062 being O-rich and the remaining four
targets having mixed envelope chemistries with both O-rich and C-rich
features (Cerrigone et al. 2009). These authors find that 40 per cent
of their sample of hot post-AGB stars have mixed chemistry, compared
with an expected fraction of less than 10 per cent, and link this to
the presence of a dusty torus which renders these objects bright in
the mid-IR and therefore more likely to appear in IR-selected
samples. A correlation between mixed chemistry and a torus is noted in
Galactic bulge PNe by Guzman-Ramirez et al. (2011) who conclude that
in these objects the mixed chemistry arises due to UV irradiation of a
dense torus. The mixed chemistry objects in our sample also have
morphologies in H$_2$ or Br$\gamma$ which are consistent with the
presence of a torus and in I18379 and I22023 we also see velocity
structure in the H$_2$ lines consistent with bipolar outflow.

I19200, however, is located in a region of the $J-K$ versus $K-[25]$
diagram occupied by cooler G and F-type SOLE pre-PNe (Fig. 11), having
a bluer $K-[25]$ colour than our other targets. The $25~\umu$m flux is
4 times weaker than that of I19336 and 10 times weaker than that of
I18379 and I22023. This suggests that I19200 is less dusty, and has
ejected a lower-mass envelope than the other objects. This is
supported by modelling of the thermal dust emission which gives a
total envelope mass of $7\times 10^{-4}$~M$_\odot$ (Cerrigone et
al. 2009). The weak mid-IR continuum from I19200 makes it unlikely
that this object harbours a dusty torus, which would be consistent
with its C-rich nature. In a {\em Spitzer} survey of 140 young
Galactic PNe, Stanghellini et al. (2012) find that C-rich PNe tend to
be more symmetric (i.e. round or elliptical) than O-rich or mixed
chemistry nebulae.

The H$_2$ emission in I19200 appears as a ring around the stellar
position with a radius of 0.8 arcsec. As noted in Sec.~3.3, the ring
bulges out slightly to the SW. The H$_2$ quadrupole transitions are
optically thin and so trace the distribution of molecular
material. The simplest interpretation is that the H$_2$ emission
region forms a shell around the star and may have a mildly ovoid
geometry.

It is possible that I19200 has a more axisymmetric structure, such as
an hourglass or cylinder, and is viewed pole-on. The bulge PN
G357.2+02.0 may be such an object, appearing perfectly ring-like in
H$\alpha$ images but having a velocity profile in [O~{\sevensize III}]
which suggests a more complicated geometry (Gesicki et al. 2014); this
object also has mixed chemistry (Stanghellini et al. 2012). The C-rich
nature of I19200, its weak mid-IR flux when compared with our other
targets which do show axisymmetry, and the circular projection of the
H$_2$ emission on the sky collectively lead us to believe that this
object most likely has a shell-like envelope. We do not detect a
variation in radial velocity across the shell in the 1--0~S(1) line
with an upper limit of 15~km~s$^{-1}$.  In addition, the velocities
obtained from the 1--0~S(1) and Br$\gamma$ lines agree within error
(Table~3), even though these lines are emitted on very different
spatial scales, which suggests that they are systemic velocities.

I19200 may be evolving into a ``round'' PN, a rare morphological group
comprising less than 4 per cent of PNe in the survey of Sahai et
al. (2011). These objects may arise from the post-AGB evolution of
single stars which, lacking a binary companion, do not develop fast
collimated winds and instead expel a spherical, low mass shell (Soker
2002). So far no round pre-PNe have been discovered (Sahai et
al. 2011, 2007), most likely because their dust envelopes are too
tenuous to be detected by scattered light.

\subsection{Evolutionary status of the objects}

I19200 is the only object in our sample without a He~{\sevensize I}
detection, so that it is not yet hot enough to ionize He and is
therefore likely to be the least evolved object.  Arkhipova et
al. (2004) detect strong but variable H$\alpha$ and H$\beta$ emission
between 2001 and 2003 and classify the star as early B-type. H$\alpha$
was also detected in 2003 by Emprechtinger et al. (2005). However
Gauba et al. (2003) did not detect H$\alpha$ in their 2000
observations and Suarez et al. (2006) classify the star as Fe.  The
hydrogen densities and ionized masses determined from the Br$\gamma$
emission (Table 4) are also consistent with the recent onset of
photoionization.

I18062 and I18379 are very young PNe. Both objects show Br$\gamma$ and
He~{\sevensize I} emission indicating ionized H and He, but their
photoionized regions are still unresolved in our observations and have
densities $n_{\rm H}\sim 10^{5}$~cm$^{-3}$. I18062 appears to have
evolved rapidly in spectral type over the last few decades, with a
linearly increasing Br$\gamma$ and radio flux pointing to the onset of
ionization around 1990, so that we are witnessing the birth of this
PN. Both objects also have an axisymmetric structure to their
molecular envelopes due to previous shaping processes occurring in the
pre-PN phase. The H$_2$ emission peaks seen in I18062 resemble an
edge-on torus and in this respect it is similar to the B-type very
early PN IRAS~19306+1407 (Lowe \& Gledhill 2006). Pre-PN objects such
as IRAS~17436+5003, IRAS~06530-0213 and IRAS~19374+2359 have similar
structure, with dust peaks to either side of the star, visible in the
near-IR (Gledhill et al. 2001, Gledhill 2005), indicating a modest
degree of axisymmetry in the envelope. We suggest that as the ionized
region expands into and dissociates the molecular envelope, I18062
will become an elliptical PN. I18379 also has an elliptical structure
in H$_2$ emission, with an edge-brightened shell with the NE rim
approaching us and the SW rim receding.

I22023 and I20462 are at a more advanced phase of evolution, where the
region of ionized hydrogen has expanded into the molecular envelope.
I22023 is the most morphologically complex object in our sample with
distinct knots of H$_2$ emission arranged around the star; there is no
obvious single axis although we see a distinct velocity gradient of
$\sim 40$~km~s$^{-1}$ E-W across the object. The Br$\gamma$ emission
shows a different structure with a central ring and protrusions along
PA 100~$\deg$ and this in turn is offset from the major axis of the
8.4~GHz emission. In I20462 the molecular material appears to have
been mostly dissociated, with just a localised blob of H$_2$ emission
left against a constant low-level background. The Br$\gamma$ nebula forms
an edge-brightened ring which sits within the optical dust-scattering
nebula seen in $V$-band {\em HST} images (Ueta et al. 2000). Both
I22023 and I20462 show blue-shifted absorption in the $2.058~\umu$m
He~{\sevensize I} line. 

I19336 has the most extended Br$\gamma$ nebula in our sample, a
resolved torus of He~{\sevensize I} emssion, and appears to be at the
most advanced stage of PN formation. The H$_2$ emission forms a weak
background across the field but with a similar radial velocity (within
errors) to that of the Br$\gamma$ and He~{\sevensize I} emission, so
that we believe it to be associated with the object. The Br$\gamma$
emission shows distinct collimated lobes extending from a ring with
peaks on either side of the star. In contrast, the He~{\sevensize I}
emission forms a smaller central ring sitting within, but with similar
geometry to, the Br$\gamma$ ring. These structures give two snapshots
in time showing the propagation of the ionization fronts in this young
PN.  When the star was cooler, the Br$\gamma$ emission presumably also
formed a central ring but the H ionization front has since expanded
and been shaped by the molecular gas which it has now dissociated.
The He ionization front is now tracing the same path followed by the H
ionization front. The total mass and the mass of ionized gas estimated
from dust models, radio flux and Br$\gamma$ emission (Cerrigone et
al. 2009, 2008 and this work, respectively) are of the same order, so
that nearly all of the H in this object appears to be ionized.

\section{Conclusions}

Integral field spectroscopy in the $K$ band is a powerful tool for the
investigation of hot post-AGB objects which are about to become
PNe. This wavelength range contains a rich spectrum of ro-vibrational
H$_2$ lines, fluoresced by the increasing UV flux from the central
star, and revealing the distribution of molecular material in these
objects. The line ratios highlight regions of the objects where the
H$_2$ excitation deviates from pure UV pumping and includes a thermal
contribution, likely due to UV heating in dense gas.

The developing photoionization regions immediately surrounding the
stars are traced by hydrogen Br$\gamma$, and for hotter stars
He~{\sevensize I} $2.058~\mu$m lines.  When combined with AO
correction on an 8~m telescope such as Gemini the line emission and
continuum can be imaged with a spatial resolution of better than
$0.2$~arcsec, which can be improved further with image deconvolution.

We have applied the technique to six hot post-AGB stars. All six
objects show Br$\gamma$ emission and evidence for photoionization,
with I19336, I20462 and I22023 having extended Br$\gamma$
nebulae. Although the distances are uncertain, it seems likely that
I19336, I20462 and I22023 are more evolved than I18062, I18379 and
I19200, where the ionized regions are still unresolved.

I19336 appears to be the most evolved object, with a bipolar
H~{\sevensize II} region extending beyond our NIFS field of view; the
He$^{+}$ region in this object is also resolved as a ring-like
structure revealed by the He~{\sevensize I}~$2.058~\umu$m line. The
absence of any strong H$_2$ emission suggests that molecular material
has been dissociated over the inner volume of the emission line
nebula. This is also the case in I20462, although a clump of emitting
H$_2$ remains in this object.

The H$_2$ emission in I19200 forms a ring around the star.  It has a
weak 25~$\umu$m flux compared with the other targets suggesting a low
dust mass and we do not detect any strong radial velocity signatures.
We argue that this C-rich object has a shell-like envelope and may be
about to evolve into a ``round'' PN.
 
I18062 appears to be a rapidly evolving object. The linear increase in
the Br$\gamma$ and 8.4~GHz fluxes between 1999 and 2007 implies that
the ionization began recently, around 1990. The H$_2$ emission is
suggestive of a toroidal distribution of material, with faint
orthogonal bipolar entensions. The developing H~{\sevensize II} region
is unresolved at present, but we may expect it to expand and
eventually encounter the molecular material, at which point it may
begin to resemble I19336, as the molecular torus is gradually
dissociated and then ionized, forming a bipolar emission line nebula
with equatorial desnity enhancements.  Objects such as these should
give important clues as to how the exapnding photoionized regions
interact with molecular material expelled and shaped during the pre-PN
phase, leading to the formation of PNe.

In contrast to the extended and structured line emission from
molecular and atomic material, we do not find any strong evidence for
dust-scattered continnum in these objects. In Section~4.4 we {\bf
  compare} these embryonic PNe with the pre-PN SOLE objects of Ueta
et al. (2000), which are characterised by faint reflection nebulae
surrounding bright stars.  We suggest that the objects in our sample
would have been SOLE type pre-PNe, before their stars began to ionize
the envelopes.\\

\section*{Acknowledgments}
We thank the anonymous referee whose valuable comments have
significantly improved the paper.  We thank James Hilling for
assistance with the initial processing of the observations.  Based on
observations obtained at the Gemini Observatory, which is operated by
the Association of Universities for Research in Astronomy, Inc., under
a cooperative agreement with the NSF on behalf of the Gemini
partnership: the National Science Foundation (United States), the
National Research Council (Canada), CONICYT (Chile), the Australian
Research Council (Australia), Minist\'{e}rio da Ci\^{e}ncia,
Tecnologia e Inova\c{c}\~{a}o (Brazil) and Ministerio de Ciencia,
Tecnolog\'{i}a e Innovaci\'{o}n Productiva (Argentina).

\appendix
\section{Ionized mass}
Kwok (2000) give an expression for the Br$\gamma$ flux from an optically thin nebula of
radius $R$ and distance $d$ (their equation 3.25):
\begin{equation}
F=3.41\times 10^{-27}\left(\frac{R^{3}\epsilon}{3d^{2}}\right)n_{\rm e}n_{\rm H}~{\rm erg~cm^{-2}~s^{-1}}
\end{equation}
where $\epsilon$ is the filling factor and $n_{\rm H}$ and $n_{\rm e}$ are the hydrogen (proton) and electron densities. This expression can be used to obtain a relationship between the observed angular 
radius ($\phi$) and line flux of the nebula, and the distance and density, giving:
\begin{equation}
\phi=292.5\left(\frac{F_{\rm W-17}}{d_{\rm kpc}\epsilon n_{\rm e}n_{\rm H}}\right)^{1/3}~{\rm arcsec}
\end{equation}
where $F_{\rm W-17}$ is the Br$\gamma$ flux in units of $10^{-17}$~W~m$^{-2}$ and $d_{\rm kpc}$
is the distance in kpc. For a He/H fraction of 0.11 (He abundance of 0.1) and assuming that
50 per cent of the He is ionized, then $n_{\rm e}=1.055 n_{\rm H}$. Inserting into Equation 2
and rearranging gives an expression for the density in cm$^{-3}$:
\begin{equation}
n_{\rm H}=4.87\times 10^{3}F_{\rm W-17}^{0.5}\epsilon^{-0.5}d_{\rm kpc}^{-0.5}\phi^{-1.5}~{\rm cm^{-3}}
\end{equation}
The ionized mass in the nebula is given by (Kwok 2000 equation 4.36):
\begin{equation}
M_{\rm i}=\frac{4\pi}{3}n_{\rm H}\mu m_{\rm H}\epsilon R^{3}
\end{equation}
where $m_{\rm H}$ is the mass of a hydrogen atom and $\mu$ (the mean atomic weight per proton)
is 1.44 for a He fraction of 0.11. Substitution of (3) into (4), replacing $R$ by the
angular radius $\phi$ and converting to solar masses gives:
\begin{equation}
M_{\rm i}=8.26\times 10^{-5}F_{\rm W-17}^{0.5}\epsilon^{0.5}d_{\rm kpc}^{2.5}\phi^{1.5}~{\rm M_\odot}
\end{equation}
We assume that $\epsilon=0.6$ corresponding to a shell of thickness $\Delta R = 0.26R$.

\section{H$_2$ line fluxes and peak wavelengths}

\begin{table*}
\caption{H$_2$ line fluxes (in 10$^{-17}$~W~m$^{-2} \pm 0.01$) and peak wavelengths (in
$\umu$m~$\pm 0.0001$).}
\begin{tabular}{lcccccccccc}
Line & \multicolumn{2}{c}{I18062} & \multicolumn{2}{c}{I18379} & \multicolumn{2}{c}{I19200} & \multicolumn{2}{c}{I20462} & \multicolumn{2}{c}{I22023}  \\ 
     & F  & $\lambda$ & F  & $\lambda$ & F  & $\lambda$ & F  & $\lambda$ & F    & $\lambda$ \\
\hline \\
7-5 O(5) &0.03&2.0226&0.21&2.0213&0.22&2.0219&  - &  -   &0.23&2.0213    \\
6-4 O(7) & -  &  -   &  - &  -   &0.08&2.0296&  - &  -   &  - &  -       \\
1-0 S(2) &0.78&2.0343&2.13&2.0329&1.76&2.0335&0.07&2.0336&1.65&2.0328    \\
8-6 O(3) &0.14&2.0423&0.29&2.0407&0.57&2.0415&  - &  -   &0.46&2.0408    \\
3-2 S(5) & -  &  -   &0.15&2.0645&0.13&2.0652&  - &  -   &0.10&2.0647    \\
2-1 S(3) &0.32&2.0737&0.98&2.0722&1.24&2.0731&  - &  -   &1.45&2.0722    \\
9-7 Q(2) &0.03&2.0844&  - &  -   &0.05&2.0837&  - &  -   &0.12&2.0830    \\
9-7 Q(3) &0.04&2.1008&0.09&2.0995&0.11&2.1002&  - &  -   &0.25&2.0995    \\
7-5 O(6) & -  &  -   &  - &  -   &0.06&2.1085&  - &  -   &0.11&2.1077    \\
1-0 S(1) &3.09&2.1219&6.26&2.1206&6.53&2.1213&0.15&2.1213&4.80&2.1205    \\
3-2 S(4) &0.04&2.1281&0.18&2.1268&0.16&2.1275&  - &  -   &0.15&2.1266    \\
4-3 S(6) & -  &  -   &  - &  -   &0.05&2.1513&  - &  -   &  - &  -       \\
2-1 S(2) &0.14&2.1544&0.63&2.1530&0.50&2.1536&  - &  -   &0.63&2.1529    \\
9-7 O(2) &0.02&2.1729&  - &  -   &0.06&2.1721&  - &  -   &0.13&2.1713    \\
3-2 S(3) &0.16&2.2015&0.50&2.2000&0.47&2.2007&0.02&2.2006&0.67&2.2000    \\
4-3 S(5) & -  &  -   & -  &  -   &0.05&2.2044&  - &  -   &0.13&2.2036    \\
8-6 O(5) &0.06&2.2109&0.16&2.2094&0.10&2.2101&  - &  -   &0.35&2.2094    \\
1-0 S(0) &0.78&2.2234&1.78&2.2220&1.68&2.2227&0.06&2.2226&1.56&2.2218    \\
2-1 S(1) &0.51&2.2478&1.16&2.2463&1.16&2.2471&0.03&2.2469&1.41&2.2463    \\
9-7 O(3) &0.08&2.2538&0.17&2.2523&0.23&2.2532&  - &  -   &0.32&2.2523    \\
4-3 S(4) &0.02&2.2669& -  &  -   &0.05&2.2662&  - &  -   &0.10&2.2654    \\
3-2 S(2) &0.10&2.2873&0.23&2.2858&0.20&2.2865&  - &  -   &0.41&2.2857    \\
10-8 Q(1)&0.02&2.3227&0.04&2.3213&0.05&2.3220&  - &  -   &0.10&2.3211    \\
4-3 S(3) &0.07&2.3447&0.17&2.3434&0.19&2.3441&  - &  -   &0.33&2.3432    \\
9-7 O(4) &0.03&2.3460&0.08&2.3444&0.05&2.3452&  - &  -   &0.15&2.3445    \\
2-1 S(0) &0.14&2.3560&0.48&2.3545&0.38&2.3553&  - &  -   &0.72&2.3544    \\
3-2 S(1) &0.16&2.3869&0.58&2.3854&0.39&2.3862&  - &  -   &0.92&2.3854    \\
1-0 Q(1) &2.37&2.4071&6.24&2.4057&5.69&2.4065&0.20&2.4064&4.73&2.4056    \\
1-0 Q(2) &0.83&2.4141&2.04&2.4125&1.84&2.4134&0.06&2.4134&1.89&2.4125    \\
1-0 Q(3) &1.47&2.4245&4.13&2.4228&3.99&2.4238&0.12&2.4236&3.62&2.4229    \\
\hline \\
\end{tabular}
\end{table*}

\end{document}